%% file: main.tex
\documentclass[
  journal=pasa,
  manuscript=research-paper, %% or "review"
  year=2021,
  volume=37,
]{cup-journal}

\usepackage{natbib}
\usepackage{microtype,siunitx,booktabs}
\newcommand{\gh}{GLEAM\,\-J0917\-$-$0012}
\newcommand{\red}[1]{\textcolor{black}{#1}}
\title[Grism Observations of the Candidate uHzRG \gh]{{\it HST} WFC3/Grism Observations of the Candidate Ultra-High-Redshift Radio Galaxy \gh}

\author{N. Seymour}
\affiliation{International Centre for Radio Astronomy Research, Curtin University, 1 Turner Avenue, Bentley, WA 6102, Australia}
\email[N. Seymour]{nick.seymour@curtin.edu.au}
\author{G. Drouart}
\affiliation{International Centre for Radio Astronomy Research, Curtin University, 1 Turner Avenue, Bentley, WA 6102, Australia}
\author{G. Noirot}
\affiliation{Institute for Computational Astrophysics and Department of Astronomy \& Physics, Saint Mary's University, 923 Robie Street, Halifax, NS B3H 3C3, Canada}
\author{J.W. Broderick}
\affiliation{International Centre for Radio Astronomy Research, Curtin University, 1 Turner Avenue, Bentley, WA 6102, Australia}
\author{R.J. Turner}
\affiliation{School of Natural Sciences, University of Tasmania, Private Bag 37, Hobart, 7001, Australia}
\author{S.S. Shabala}
\affiliation{School of Natural Sciences, University of Tasmania, Private Bag 37, Hobart, 7001, Australia}
\author{D.K. Stern}
\affiliation{Jet Propulsion Laboratory, California Institute of Technology, 4800 Oak Grove Drive, Pasadena, CA 91109, USA}
\author{S. Bellstedt}
\affiliation{International Centre for Radio Astronomy Research, The University of Western Australia, 7 Fairway, Crawley, WA 6009, Australia}
\author{S. Driver}
\affiliation{International Centre for Radio Astronomy Research, The University of Western Australia, 7 Fairway, Crawley, WA 6009, Australia}
\author{L. Davies}
\affiliation{International Centre for Radio Astronomy Research, The University of Western Australia, 7 Fairway, Crawley, WA 6009, Australia}
\author{C.A. De Breuck}
\affiliation{European Southern Observatory, Karl Schwarzschild Strasse, D-85748 Garching bei M\"unchen, Germany}
\author{J. Afonso}
\affiliation{Instituto de Astrof\'isica e Ci\^encias do Espaço, Faculdade de Ci\^encias, Universidade de Lisboa, OAL, Tapada da Ajuda, PT1349-018 Lisboa, Portugal}
\affiliation{Departamento de F\'isica, Faculdade de Ci\^encias, Universidade de Lisboa, Edif\'icio C8, Campo Grande, PT1749-016 Lisbon, Portugal}
\author{J.D.R. Vernet}
\affiliation{European Southern Observatory, Karl Schwarzschild Strasse, D-85748 Garching bei M\"unchen, Germany}
\author{T.J. Galvin}
\affiliation{International Centre for Radio Astronomy Research, Curtin University, 1 Turner Avenue, Bentley, WA 6102, Australia}

\doi{10.1017/pasa.2020.32}

\received {dd Mmm YYYY}
\revised  {dd Mmm YYYY}
\accepted {dd Mmm YYYY}
\published{22 September 2020}

 %% First letter not capped
\begin{document}

\input{content}

% PASA uses footnotes, not endnotes. \endnote in this template will behave like \footnote; and \printendnotes will not output anything.
% \printendnotes

%\bibliographystyle{aas-macros}
%\bibliographystyle{pasa-mnras}
\bibliography{main}
%\bibliography{example}

%\appendix

%\input{appendix}

\end{document}

%% file: content.tex
\begin{abstract}
We present {\it Hubble Space Telescope} Wide Field Camera 3 photometric and grism observations of the candidate ultra-high-redshift ($z>7$) radio galaxy, \gh. This radio source was selected due to the curvature in its $70-230\,$MHz, low-frequency Murchison Widefield Array radio spectrum and its faintness in $K$-band. Follow-up spectroscopic observations of this source with the Jansky Very Large Array and Atacama Large Millimetre Array were inconclusive as to its redshift. Our F105W and F0986M imaging observations detect the host of \gh~and a companion galaxy, $\sim$~one arcsec away. The G102 grism observations reveal a single \red{weak} line in each of the spectra of the host and the companion. To help identify these lines we utilised several photometric redshift techniques including template fitting to the grism spectra, fitting the \red{ultraviolet (UV)}-to-radio photometry with galaxy templates plus a synchrotron model, fitting of the UV-to-near-\red{infrared (IR)} photometry with {\tt EAZY}, and fitting the radio data alone with {\tt RAiSERed}. For the host of \gh~we find a line at $1.12\,\mu$m and the UV-to-radio spectral energy distribution fitting favours solutions at $z\sim 2$ or $z\sim 8$. \red{While this fitting shows a weak preference for the lower redshift solution, the models from the higher redshift solution are more consistent with the strength of the spectral line.} The redshift constraint by {\tt RAiSERed} of $>6.5$ \red{also supports} the interpretation that this line could be Lyman$-\alpha$ at $z=8.21$\red{; however {\tt EAZY} favours the $z\sim 2$ solution. We discuss the implications of both solutions}. For the companion galaxy we find a line at $0.98\,\mu$m and the spectral energy distribution fitting favours solutions at $z<3$ implying that the line could be the [OII]$\lambda3727$ doublet at $z=1.63$ \red{(although the {\tt EAZY} solution is $z\sim 2.6\pm 0.5$)}. Further observations are still required to unambiguously determine the redshift of this intriguing candidate ultra-high-redshift radio galaxy. 
%Confirmation of a powerful radio galaxy well within the epoch of reionisation would have wide ranging implications. 
\end{abstract}

\section{INTRODUCTION}
\label{sec:intro}

Active galactic nuclei (AGN) are now found regularly in the early Universe from searches for unobscured quasi-stellar objects \citep[QSOs; e.g.][]{Banados:16}. Indeed, the most distant AGN currently known is a QSO with a redshift of $z=7.642$ \citep{Wang:21}, putting it just $\sim 670\,$Myr after the Big Bang. The super-massive black hole (SMBH) powering this AGN is estimated to have a mass of $1.6\times 10^9\,M_\odot$. Such observations naturally raise the questions of how did SMBHs form so early and/or grow so quickly? The existence of billion-solar-mass black holes in the early Universe leads to tension with current models of their formation, even with the black holes growing at their maximum Eddington rate \citep[e.g.][]{Volonteri:12,Smith:17}. Did these first SMBHs form from massive seeds? Did they form from small seeds but accrete and/or merge rapidly?  

Many AGN are obscured, meaning that the high ultraviolet\,(UV)--optical luminosity from their accretion disc is hidden by dust and hence are missed by current QSO searches in the optical and near-infrared (near-IR). If obscured and radio-loud AGN, i.e. radio galaxies, can also be found at high redshifts, then a more complete census of AGN activity, and SMBH growth, in the early Universe may be obtained. Specifically, (a) such AGN do not outshine their host galaxies in the optical/near-IR, allowing their hosts to be studied in detail, and (b) the influence on the environment from their powerful jets and inverse-Compton emission (from interaction with the cosmic microwave background) can be investigated. Furthermore, if a suitably bright, radio-loud AGN could be discovered within the epoch of reionisation (EoR; $6.5\le z\le 15$), then the distribution of neutral hydrogen (HI) along the line of sight could be measured from absorption studies of the 21-cm line \citep[e.g.][]{Carilli:04,Ciardi:15}. 

Historically (1960--2000), the most distant AGN were frequently discovered from radio surveys since the powerful radio emission acts as a beacon across cosmic time \citep[e.g.][]{Schmidt:63,Peterson:82}. These surveys found highly accreting but often obscured AGN, where the UV--optical emission from the accretion disc is hidden by a  dusty torus around the SMBH \citep{Urry:95}. The high accretion rates for obscured, type 2 AGN are only detectable via spectro-polarimetry \citep[e.g.][]{Vernet:01}, narrow emission lines \citep[e.g.][]{Nesvadba:17} or from mid-infrared observations \citep[e.g.][]{Drouart:14}. As discussed above, the most distant AGN known currently are QSOs found from deep and wide optical/near-infrared surveys using the Lyman-break technique, whereas searches in the radio are completely independent of the Lyman-break method. 

The discovery of AGN in the distant Universe from radio surveys is undergoing a renaissance due to a new generation of deep, low-frequency radio surveys such as the TIFR GMRT Sky Survey \citep[TGSS; $150\,$MHz;][]{Intema:17}, the LOFAR Two-metre Sky Survey \citep[LoTSS][]{shimwell:17}, and the GaLactic and Extragalactic All-sky Murchison Widefield Array (GLEAM) survey \citep[$70-230\,$MHz;][]{Wayth:15}. Low-frequency radio surveys provide a lever-arm to aid the selection of high-redshift radio galaxies (HzRGs; $z>5$), which likely exist in such surveys \citep[e.g.][]{Wilman:08}, from the millions of radio sources already catalogued.

Several new HzRGs have been discovered over the last few years, breaking a distance record for powerful radio galaxies that stood for nearly two decades \citep{vanBreugel:99a}\footnote{Very recently, a handful of QSOs at $z>6$ have been detected in radio surveys, e.g. at $z=6.44$ \citep{Ighina;21} and $z=6.84$ \citep{Banados:21}, although these are less radio-luminous than the HzRGs targeted here.}. \cite{Saxena:18a} used the 1.4-GHz Faint Images of the Radio Sky at Twenty-cm (FIRST) survey \citep{Becker:95} and the National Radio Astronomical Observatory (NRAO) Very Large Array (VLA) Sky Survey \citep[NVSS;][]{Condon:98} in combination with TGSS to search for uHzRGs. Using  the  ultra-steep spectrum  (USS; radio spectral index $\alpha<-1.3$\footnote{For radio spectral indices we use the convention $S_\nu\propto\nu^\alpha$.})  technique, they found a radio galaxy, TGSS\,J1530$+$1049, at $z=5.72$ \citep{Saxena:18b}. 

Recently, we have taken advantage of a different selection technique, exploiting the 20-band radio photometry provided by the first extra-galactic data release of GLEAM \citep[][]{NHW:17}. We searched for curvature in the radio spectral energy distributions (SEDs) of candidate HzRGs, which may be caused by lower-frequency turn-overs due to synchrotron self-absorption (SSA) or free--free absorption (FFA). With this technique, and searching for sources without near-infrared $K_{\rm s}$-band detections in the Visible and Infrared Survey Telescope for Astronomy Kilo-degree Infrared Galaxy (VIKING) survey \citep{Edge:13}, we identified four candidate HzRGs \cite[][hereafter D20]{Drouart:20} in the Galaxy And Mass Assembly \citep[GAMA][]{driver:11} $9^{\rm h}$ field. Using deep $K_{\rm s}$-band imaging from the HAWKI instrument \citep{Kissler-Patig:08} on the Very Large Telescope \red{(VLT)} to locate the host galaxy, we obtained $87-117$\,GHz spectra (and imaging) from the Atacama Large Millimetre Array (ALMA).  From the detection of carbon monoxide (CO) lines, we found one source, GLEAM\,J0856$+$0223, to be a powerful radio galaxy at $z=5.55$ (D20), more than five times as luminous at low frequencies than TGSS\,J1530$+$1049 at a similar redshift. 

Furthermore, we found a \red{candidate} {\em ultra-HzRG} (uHzRG, $z>7$), \gh, potentially at $z\sim 10$ based on \red{three} weak CO features \red{(with signal-to-noise ratios, SNR, of $2.8-3.4$, although a joint fit found a collective significance of $\sim 5.4\sigma$)}. Follow-up spectroscopic observations with both ALMA (targeting the same CO lines) and with the VLA (targeting lower transition lines around $42\,$GHz) were unable to confirm this possibility \citep[][hereafter D21]{Drouart:21}. Detailed analysis of the optical-to-far-IR SED, including the 100-GHz datum, suggestted that the host of \gh~lies at either $z\sim 3$ (optical $z_{phot}\sim 2.2$) or $z\sim 8$. We also identified a `companion' source about one arcsecond to the west of \gh.

In order to elucidate the nature of \gh~we have obtained spectroscopic and photometric follow-up with the {\it Hubble Space Telescope (HST)}. We report the results of these Wide Field Camera 3 (WFC3) observations here. As confirming the redshift of a galaxy in the early Universe is a complex endeavour we supplement the results of the WFC3 data with several detailed SED redshift fitting methods. The WFC3 observations and their reduction are reported in \S\ref{sec:obs}. Other data and photometry used in this work are presented in \S\ref{sec:other}. We present the SED fitting methods in \S\ref{sec:fitting}. The results of our analysis are set forth and discussed in \S\ref{sec:results}. We conclude this paper in \S\ref{sec:con}. In this paper, we use a flat $\Lambda$CDM cosmology with H$_0$=67.7 km\,s$^{-1}$\,Mpc$^{-1}$ and $\Omega_{\rm M}$=0.308 \citep{Planck:16}. All coordinates are in J2000 and all magnitudes in the AB system. 

\section{HST Observations}
\label{sec:obs}

\begin{figure*}[t]
\vskip -0.9cm
\begin{center}
\includegraphics[scale=0.75]{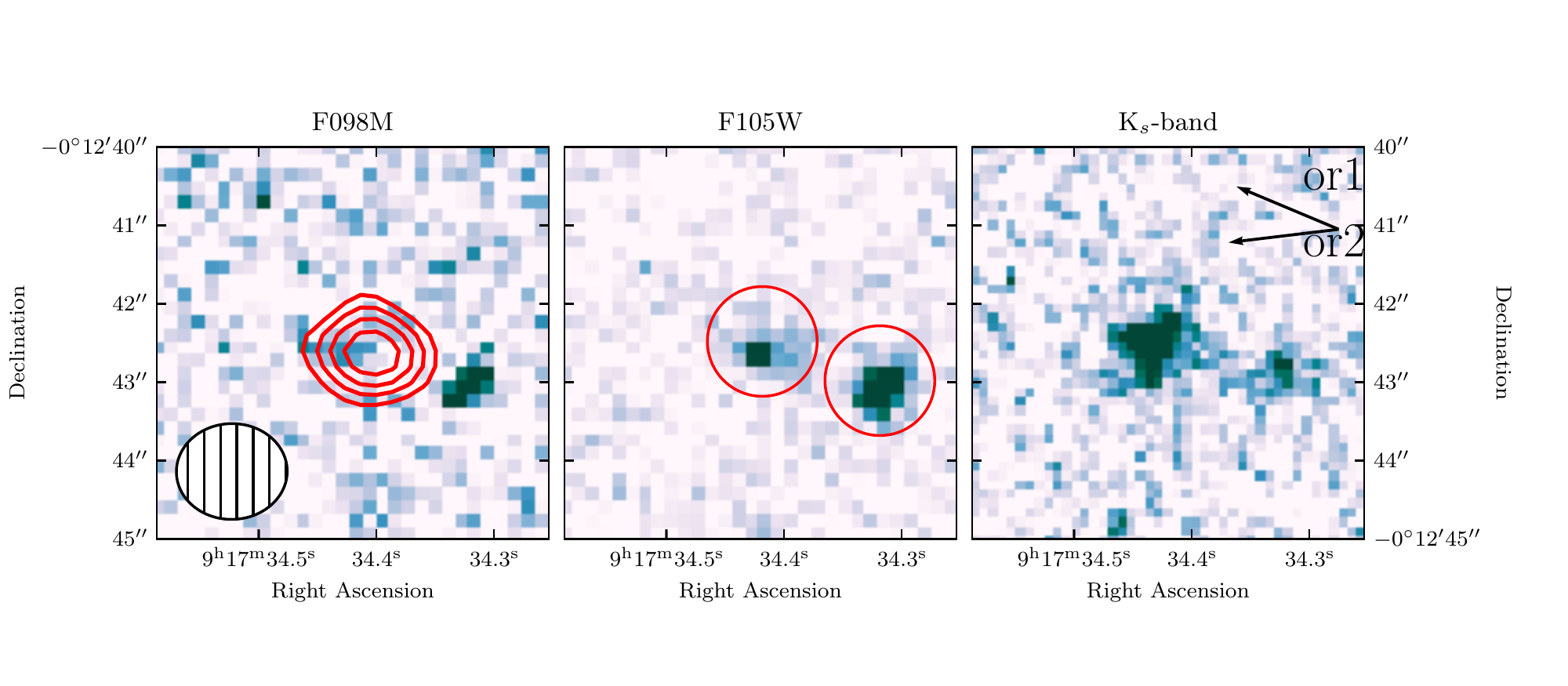}
\vskip -0.9cm
\caption{Grey-scale \red{F098M,} F105W \red{and $K_s$-band} image\red{s} of the host of \gh~\red{(centre of each panel)} and \red{the} companion galaxy \red{(to the west)}. Overlaid \red{on the F098M image} are contours from the highest-resolution radio data (100-GHz ALMA, red) with the beam shown in the lower left. The contours are $3,4,5,6\,\times\,10\,\mu$Jy\,beam$^{-1}$, the \red{root mean square (RMS) noise} of the ALMA image. The host galaxy shows a compact morphology with a size of $\sim 0.25\,$arcsec, and with a faint extension to the west. \red{The two circles in the central panel indicate the $0.7''$ radius apertures used to measure the flux densities, with the {\tt Source Extractor} positions given in Table~\ref{tab:pos}.} The two dispersion directions of each orientation of the grism are indicated by the arrows \red{in the third panel}.}
\label{fig:F105W}
\end{center}
\end{figure*}

\gh~was observed with {\it HST} under proposal ID 16184 using WFC3 for five orbits in late 2020. Considering the earlier indications that this radio galaxy could lie at $z\sim 10$ (D20), the first orbit was imaging with the F105W filter (which would have been completely blue-ward of Lyman-$\alpha$ at that redshift). However, the host of \gh~was clearly detected suggesting a lower redshift. Therefore the subsequent four orbits of grism observations were conducted with the F098M/G102 combination of filter/grism (rather than the F160W/G141 combination as originally proposed) to target the observed-frame Lyman$-\alpha$. The observations were carried out in two orientations of two orbits each. Unfortunately, one of the orbits was approximately colinear with the galaxy-companion direction which results in over-lapping of their spectral features. The dispersion directions are indicated in Fig.~\ref{fig:F105W}. We processed the data using two independent software packages to ensure any significant spectral features are not an artefact of any one code.

\subsection{Reduction with {\tt hstaxe}}
\label{sec:hstaxe}

\subsubsection{Imaging}
\label{sec:axeim}
We first used the {\tt hstaxe} v1.0.0\footnote{Initially we used the Beta 3 release, which provided the same results.} \citep{Sosey:13} software to process the data. This code is a Python and C successor to the {\tt iraf} based {\tt axe} code used for many years on {\it HST} data. We based our reduction on the cookbook presented in the Jupyter notebook available with {\tt HSTAXE}.  We used the default values in this notebook unless otherwise stated. The pipeline starts from the `flt' fits files downloaded from MAST\footnote{Barbara A. Mikulski Archive for Space Telescopes: \url{https://mast.stsci.edu}}. The {\tt AstroDrizzle} routine was used to \red{combine all the G102 grism observations to define a master pixel grid.} The `full-depth' F098M and F105W images were also created with {\tt AstroDrizzle} from all available exposures \red{using this master pixel grid} . We used the {\tt combine\_type=`median'} and {\tt combine\_nhigh=1} as recommended for more than three exposures. The pixel size of the `drizzled' image matches that of the WFC3 detector ($\sim 0.13\,$arcsec) by default. 

To detect and measure the flux densities of the host and companion, {\tt Source Extractor} \citep[][]{Bertin:96} was run on both images. \red{We used $0.7\,$arcsec radius apertures to measure the fluxes and applied an aperture correction (see \S~\ref{sec:other} for full details of the  photometry). We used the detections in the F105W band (where the host galaxy has the highest signal-to-noise ratio, SNR) to define the centres of these apertures.} The only \red{other} difference from the {\tt HSTAXE} cookbook {\tt Source Extractor} parameters was that we used a lower \texttt{DETECT\_MINAREA} of 4 (rather than the default of 10), and lower values of the parameters {\tt DETECT}{\tt \_THRESHOLD} and
\noindent
{\tt\-ANALYSIS}{\tt\-\_THRESH} (1.5 rather than the default of 3). We report the {\tt Source Extractor} positions of both galaxies in Table~\ref{tab:pos}.

\begin{table}[t]
\begin{threeparttable}
\caption{F105W~{\tt Source Extractor} coordinates of the host of \gh~and the companion.} \label{tab:pos}
\begin{tabular}{@{}lcc@{}} \toprule
Galaxy & R.A. & Declination \\ \midrule
\gh & 09:17:34.42 & $-$00:12:42.5 \\
Companion           & 09:17:34.32 & $-$00:12:43.0 \\
\bottomrule
\end{tabular}
\end{threeparttable}

\end{table}

\subsubsection{Grism Spectra}
We used the {\tt aXe} Fluxcube extraction method to create the 1D and 2D spectra. This method takes advantage of the  mosaics at different wavelengths to aid in the flux calibration and spectral extraction. We used the following steps for both orientations separately and for all the data combined:

\begin{figure*}[t]
\begin{center}
\includegraphics[width=45pc, height=15pc]{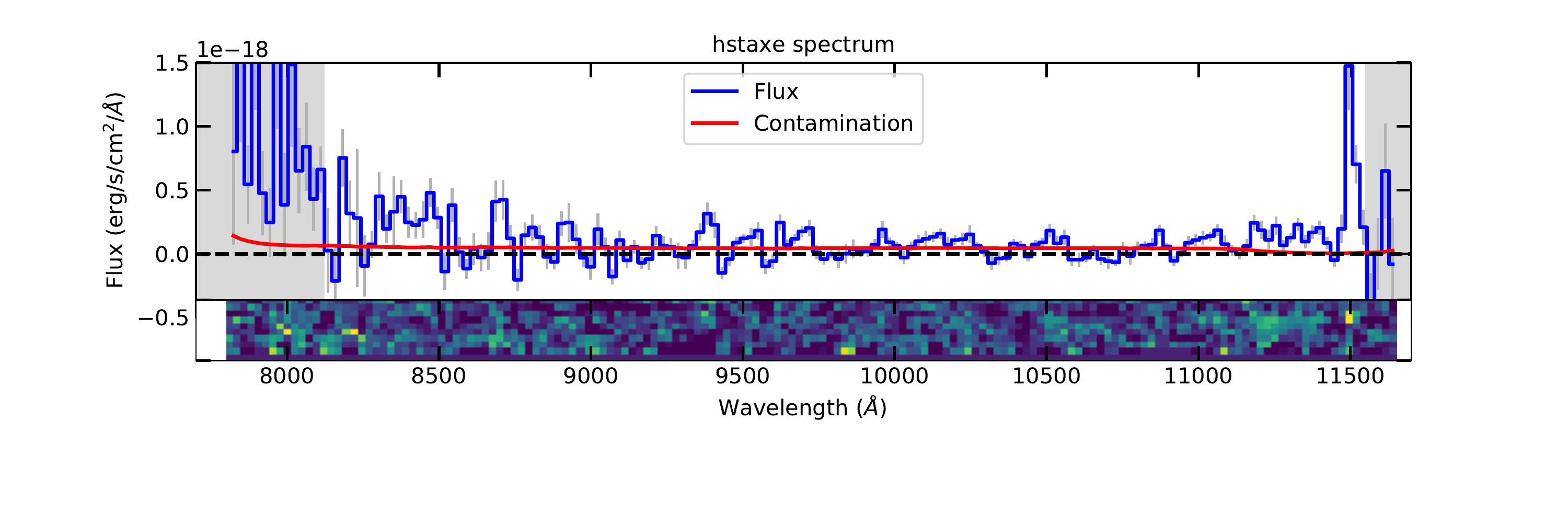}
\vskip -1.4cm
\includegraphics[width=45pc, height=12pc]{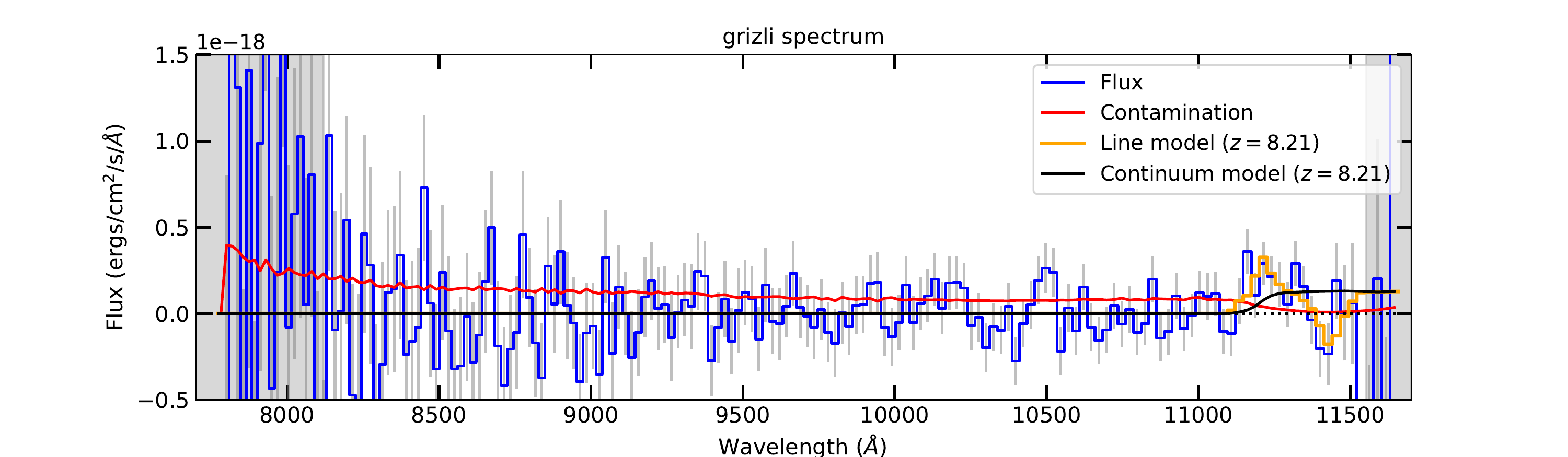}
\vskip -0.5cm
\caption{The {\tt hstaxe} 1D and 2D (top), and {\tt grizli} 1D (bottom) spectra of \gh. These are the combined spectra from both orientations. In each orientation the width of the extraction is twice the size of the source (perpendicular to the trace) of the target in the F140W image. Both reductions show the estimated contamination from other sources (indicated in red), which is subtracted from the 1D and 2D flux. This contamination occurs from the companion galaxy and only affects orientation 1. Grey shaded regions indicate where the transmission of the G102 grism drops below $10\%$. The spectrum is consistent with a very low or zero flux across most of the wavelength range. The {\tt hstaxe} 1D spectrum shows a strong feature at $1.15\,\mu$m that is not seen in the {\tt grizli} reduction which seems to an artefact. There is a fainter, but broader, feature at $1.12\,\mu$m seen in both reductions (and found with the {\tt grizli} line finding routine\red{, albeit at a SNR of $\sim 3$}). The {\tt grizli} PDF from the template fitting (Fig.~\ref{fig:grizlizpdf}) shows numerous maxima. We overlay the continuum (black) and line (orange) model template which is most consistent with our other SED constraints (see \S\ref{sec:fitting}). This solution at $z=8.21$ shows the Lyman-$\alpha$ line in emission and the NV$\lambda\lambda 1238,1242$ line in absorption. We discuss what these features in detail in \S~\ref{sec:results}.}
\label{fig:0917spec}
\end{center}
\end{figure*}

\begin{figure*}[!ht]
\begin{center}
\includegraphics[width=45pc, height=15pc]{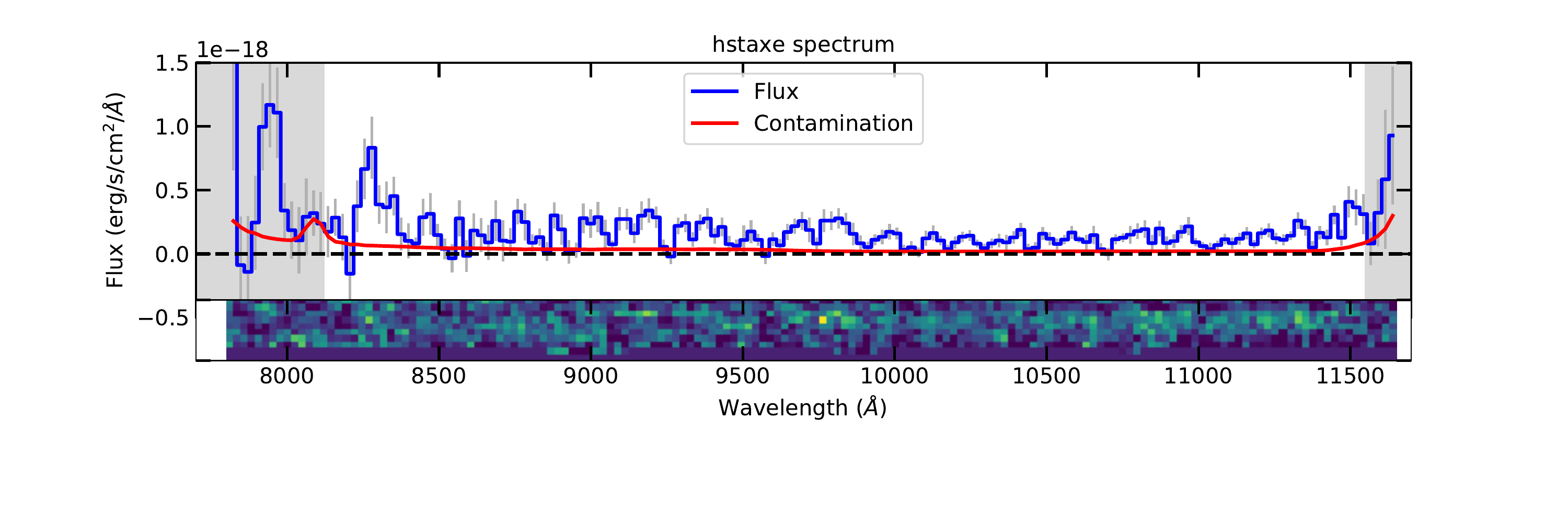}
\vskip -1.4cm
\includegraphics[width=45pc, height=12pc]{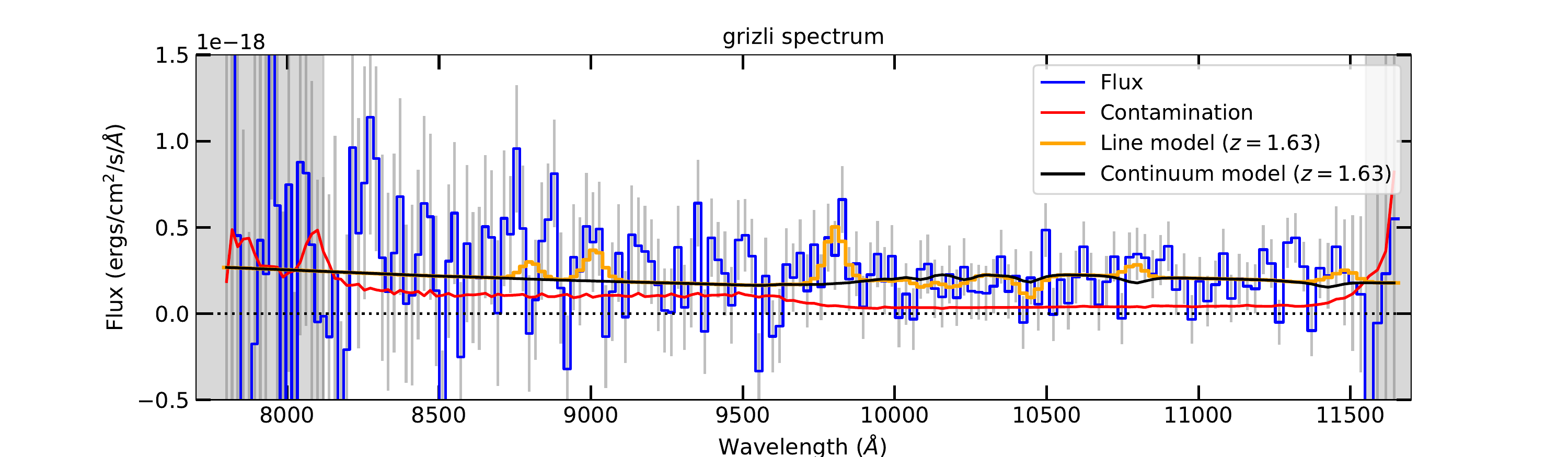}
\vskip -0.5cm
\caption{As Fig~\ref{fig:0917spec}, but for the companion galaxy. The {\tt grizli} PDF from the template fitting (Fig~\ref{fig:grizlizpdf}) shows numerous maxima. We overlay the continuum (black) and line (orange) model template which is most consistent with our other SED constraints (see \S\ref{sec:fitting}). This solution is $z=1.63$ with the central line at $0.98\,\mu$m identified as the [OII]$\lambda 3727$ doublet. We discuss the likelihood of this solution in conjunction with the SED fitting in \S~\ref{sec:results}.}
\label{fig:comspec}
\end{center}
\end{figure*}

\begin{enumerate}
    \item We used the {\tt iolprep aXe} task to generate separate F098M catalogues derived from the full-depth catalogue and matched to the geometry of each F098M pre-imaging exposure (note the full-depth catalogue is created with the F105W image as a detection image). 

    \item We created a file, {\tt aXe.lis}, with each row containing a G102 grism exposure name, the expected F098M catalogue names (created in the previous step) and the name of the associated F098M direct-preimage file.

    \item We created a {\tt cubelist.lis} file with each row containing the name of the full-depth images, the pivot wavelength of each filter (i.e. PHOTPLAM in nm), and the AB magnitude zero-points.

    \item We then created the `flexcube' model (with {\tt axetasks.\-fcubeprep}), which is a combination of the mosaics and segmentation file into a master `FLX' file for each full-depth image and an `FLX' file for each G102 observation. These files can be used to compute the contamination of spectra and to perform the extraction.

    \item We ran {\tt aXeprep} on the data, which takes the {\tt aXe.lis} file, the default grism configuration file (in this case G102.\-F098M.\-V4.32.conf\footnote{\label{fnre}available from: \url{https://www.stsci.edu/hst/instrumentation/wfc3/documentation/grism-resources}}) and subtracts the background. Also specified at this stage is the {\tt mfwhm} parameter, which determines the extent of the area that is masked perpendicular to the trace of each object before the background is determined. We used {\tt mfwhm}~$=1$ (rather than the default of 2), which is better for a lower signal-to-noise ratio and as we have many sources with {\tt Source Extractor} from the lower detection threshold (hence the field is more crowded).

    \item We ran {\tt axecore} using the flexcube models, which takes the {\tt aXe.lis} and configuration files. We also specify at this step the precise method to extract the spectra. In this case we turned on the titled extraction ({\tt orient}$=$True) and slitless geometry ({\tt slitless\_geom}$=$True). We also used values of {\tt extrfwhm}~$=2$ and {\tt drzfwhm}~$=1$, smaller than the defaults. See the {\tt hstaXe} documentation for more details\footnote{\url{https://github.com/spacetelescope/hstaxe/blob/master/docs/hstaxe/axe_tasks.rst}}. 

    \item We then used {\tt drzprep} with optimal extraction to prepare the individual spectra from each G102 exposure for combining.

    \item Finally, we used the {\tt axecrr} task to perform an optimal extraction and drizzling of the 1D and 2D spectrum with the same {\tt aXe.lis} and configuration files. This task combines all the data with default cosmic ray rejection (CRR) as provided in the configuration files. We also used values of {\tt infwhm} and {\tt outfwhm} equal to the values for {\tt extrfwhm} and {\tt drzfwhm} used with {\tt axecore} (respectively). In contrast to the default pipeline, the uncertainty on each spaxel of the spectrum is determined from the RMS of the full 2D spectrum in the cross-dispersion direction then scaled to the appropriate flux by the sensitivity file: WFC3.IR.G102.1st.\-sens.2.fits\footnote{see Footnote~\ref{fnre}}.
\end{enumerate}

The final 1D and 2D {\tt hstaxe} spectra for \gh~are presented in the top panel of Fig.~\ref{fig:0917spec}. The estimated contamination (indicated in the 1D spectra) is subtracted from both and only shown for reference. The contamination is only seen in one orientation (Or1), given it is approximately coaligned with  the host galaxy/companion vector (see Fig.~\ref{fig:F105W}). The final 1D and 2D {\tt hstaxe} spectra for the companion are presented in the top panel of Fig.~\ref{fig:comspec}, likewise with the contamination subtracted. 

\subsection{Reduction with {\tt grizli}}
\label{Sec:grzili}

We also reduced the {\it HST} data using the Grism redshift \& line ({\tt grizli}) analysis software \citep{Brammer:19}. This code is designed to offer a single tool to perform all the stages from the reduction to the spectral characterization of current {\it HST} and future {\it James Webb Space Telescope (JWST)}\footnote{\url{https://www.stsci.edu/jwst}} slitless spectroscopic observations. We used {\tt Grizli} version 1.0.dev1365 and followed the standard reduction steps, contamination modelling, and spectral template fitting for redshift determination, as are outlined in the official documentation. We now provide a summary of the steps we carried out:

\begin{enumerate}
    \item Starting from the raw {\tt flt} files from MAST, we first used the {\tt grizli} function {\tt parse\_visits}, with default parameters, to associate the individual exposures into visit groups taken in the same filter and position angle. The grism and corresponding direct imaging association was also performed at this step.

    \item We then used the {\tt preprocess} function which performs the flat-fielding, pixel flagging (cosmic-ray rejection and bad pixels masking with {\tt AstroDrizzle}, and persistence masking), relative astrometric alignment, sky background subtraction, and drizzling (with {\tt AstroDrizzle}) of all the direct imaging and grism visits. We only used default {\tt grizli} parameters in this multi-step preprocessing and refer the reader to the official documentation for further details on these steps.

    \item We then performed an absolute astrometric alignment of the individual exposures to the {\it Gaia} \citep{gaia:16} data release two \citep[DR2;][]{gaia:18} with {\tt grizli}’s {\tt fine\_alignment} function and default parameters. 

    \item After the fine alignment step, we created a F098M + F105W direct image mosaic with {\tt grizli}’s {\tt drizzle\_overlaps} with a final mosaic pixel fraction of ${\tt pixfrac}=0.75$, and we also masked diffraction spikes on the mosaic image with {\tt mask\_IR\_psf\_spike} for sources brighter than 17~mag AB. We then performed source detection on the mosaic image with {\tt multiband\_catalog} which uses the {\tt Source Extractor} Python wrapper {\tt sep} (Barbary 2016). We used default {\tt grizli} parameters, including a detection threshold ({\tt threshold}) of $1\sigma$ above the global background RMS, a minimum source area in pixels {\tt minarea} of $9$, and a deblending contrast ratio {\tt deblend\_cont} and a number of deblending thresholds {\tt deblend\_ntresh} of $0.001$ and $32$, respectively. Matched-aperture photometry in the F098M and F105W bands was simultaneously performed during this step.

    \item The last step before spectral extraction is the continuum modelling of sources in the frames for contamination estimate and removal. For this step, we used {\tt grism\_prep} and {\tt multifit.GroupFLT} as recommended in the documentation, using the mosaic catalog as the source detection catalog. The continuum modelling is an iterative process done in two steps. In the first step, a linear continuum model of all sources in the field down to the default {\tt mag\_limit} of $25$~mag is created. Then, {\tt grizli} refines the models by fitting higher-order polynomials (here ${\tt poly\_order} = 3$) to all sources iteratively in descending brightness, here from 18 to 24~mag, after subtracting potential contamination from neighbouring objects. {\tt Grizli} iterates {\tt refine\_niter} times over the refined modelling, with ${\tt refine\_niter}=3$ here.

    \item We then extracted the 2D cutouts of the host of \gh~ and of the companion source with {\tt extract} and default parameters, and redshift-fitted the spectra with {\tt fitting.run\_all\_parallel} in the range $z=0.05-10$ with the default set of FSPS \citep[Flexible Stellar Population Synthesis;][]{Conroy:09,Conroy:10} and emission line templates. The advantage of the {\tt grizli} redshift-fitting routine compared to traditional redshift fitters is that it fits the spectra in the native 2D frames by modelling the 1D spectral templates into the 2D grism space, taking into account source morphology and spectral smearing of pixels that are adjacent in the direct imaging.
\end{enumerate}

{\tt Grizli} provides the full redshift probability density function (PDF) of the fitting as well as the 1D continuum and line emission models.  We show the 1D {\tt grilzi} spectra and contamination estimates in the lower panels of Figs.~\ref{fig:0917spec} \&~\ref{fig:comspec}. Overlaid are templates selected from maxima in the PDF (Fig.~\ref{fig:grizlizpdf}) which best correspond to the spectral line observed in each spectra and the other photometric constraints on the redshift (see \S\ref{sec:results}).

\begin{table}[t!]
\begin{threeparttable}
\caption{\red{Observing log of exposures with significant detections.}}\label{tab:log}
\red{
\begin{tabular}{@{}ccccrr@{}} \toprule
Facility & Instrument & Band  & Date       & $\lambda_0$ & Time \\
         &            &       & YYYY-MM-DD &  [$\mu$m]  & [s] \\
\midrule
{\it HST} & WFC3      & F098M & \renewcommand{\arraystretch}{0.75}
\begin{tabular}{@{}l@{}}
                   2020-12-22\\
                   2020-12-27\\
                 \end{tabular} &  0.98 & 2320\\
{\it HST} & WFC3      & F105W & 2020-11-16 & 1.05 & 2580\\
VLT       & HAWKI     & $K_{\rm s}$ & 2018-05-25 & 2.15  & 4325 \\
\bottomrule
\end{tabular}
}
\end{threeparttable}
\end{table}

\section{Other Data and Photometry}
\label{sec:other}

\subsection{Data}
We compiled UV, optical, near- and mid-IR photometry from the literature in addition to the \red{{\it HST } imaging presented here and the} $K_{\rm s}$-band HAWKI imaging from D20 \red{(see Table~\ref{tab:log} for observing details of those bands)}. We used the two UV bands from the {\it GALaxy Evolution eXplorer (GALEX)} Medium Imaging Survey \citep[MIS;][]{Martin:05}, the $grizy$-band images from the Hyper Suprime-Cam (HSC) Subaru Strategic Program \citep[][]{aihara:17}, the $JH$-band images\footnote{The VIKING $Z, Y$ and $K-$band data are superseded in depth by the HSC, WFC3 and HAWKI data respectively and hence are not used.} from the VIKING survey, and mid-IR images from the {\it Widefield Infrared Survey Explorer} \citep[{\it WISE};][]{Wright:10}. 

From the HSC data both the host and companion have catalogued flux densities in all five bands, although the significance of the host flux densities varies from 1.5 to 4.6$\sigma$ (\red{where $\sigma$ is the RMS noise}). In D21, based upon visual inspection of the HSC images presented in that work, we reported the host as a non-detection. Both source extraction and visual inspection of the images gave no hint of an obvious detection. Additional analysis of the images below (\S~\ref{sec:photanal}) \red{finds only weak detections ($<3\,\sigma$).} The detection reported in the HSC data release is a `child' detection of the significant detection of the companion in $i-$band only and is most likely erroneous (private communication, HSC team). 

We used reprocessed {\it GALEX}, VIKING and {\it WISE} images in a similar fashion as described in the GAMA Panchromatic Data Release \citep[GAMA PDR;][]{driver:16}. The {\it GALEX} MIS survey provided deeper images of most of the GAMA fields compared to the all-sky survey. The individual MIS frames were combined together \red{\citep[using {\tt swarp};][]{bertin:02}} to form one image for each GAMA field. \red{The resulting point spread functions (PSFs) were 4.1\,arcsec in the far-UV (FUV, $1350-1750\,$\AA) and $5.2\,$arcsec in the near-UV (NUV, $1750-2800\,$\AA) channels.} For the VIKING data, all individual frames \red{with the same filter} were rescaled onto a common zero-point magnitude (30) before being drizzled together with a $0.339\,$arcsec pixel size. Unlike the original images from GAMA PDR, these images were not convolved down to two arcsecs, but left with a mean $0.85\,$arcsec seeing. \red{The {\it WISE} data comprise four bands ($W1, W2, W3, W4$) at  3.4, 4.6, 12 and $22\,\mu$m respectively with PSFs given in Table~\ref{tab:opt_mir}. These images were also} reprocessed in a similar fashion to GAMA PDR \citep[see][for full details of this version of the GAMA PDR]{Bellstedt:21}. 

\begin{table*}[t!]
\begin{threeparttable}
\caption{\red{Photometric properties of t}he UV to near-infrared and ALMA 100-GHz images of \gh, as well as the companion \red{galaxy. We list each band along with its effective central wavelength ($\lambda_0$), the AB zero point ($ZP$), and the seeing/resolution ($\theta_{\rm FWHM}$ - note that we provide the restoring beam parameters for the ALMA image. Where relevant, we provide the $0.7\,$arcsec aperture correction and uncertainty correction factor, $C$ (see \S\ref{sec:photanal}). We then list the aperture-corrected flux densities and uncertainties, or $3\sigma$ limits, of the host of \gh~and its companion (see \S\ref{sec:other}). Uncertainties include an additional $10\%$ for the combined uncertainty in the absolute flux calibration and aperture correction.} 
}\label{tab:opt_mir}
\red{
\begin{tabular}{@{}lcccccrr@{}} \toprule
Filter & $\lambda_0$ & ZP & $\theta_{\rm FWHM}$  & Aperture & $C$  & \multicolumn{2}{c}{$F_{\nu}$ [$\mu$Jy]} \\
       & [$\mu$m]    & &  [arcsec] & Correction &  & GJ0917 & Companion \\
\midrule
FUV & $0.153$ & 18.82 & 4.1 & n/a & \ldots &  $<0.078$ & $<0.078$ \\
NUV & $0.230$ & 20.08 & 5.2 & n/a & \ldots & $<0.258$ & $<0.258$\\
\input{tables/flux}
\noindent
W1   & 3.37  & 23.18 & 5.9 & n/a  & \ldots& $<3.30$ & $<3.30$\\
W2   & 4.62  & 22.82 & 6.5 & n/a  & \ldots& $<6.1$ & $<6.1$\\
W3   & 12.1  & 23.24 & 7.0 & n/a  & \ldots& $<32.4$ & $<32.4$\\
W4   & 22.8  & 19.60 & 12.4& n/a  & \ldots& $<153$ & $<153$\\
ALMA & 3000 & \ldots & $0.702\times 0.612$ ($359^\circ$) & \ldots & \ldots & 60$\pm$13 & $<10.0$  \\
\bottomrule
\end{tabular}
}
\end{threeparttable}
\end{table*}

In this work we also used the results of the far-IR-to-radio SED presented in D21. We do not repeat all those data here, but note that we have photometry across $0.07-100\,$GHz from numerous radio facilities (described in D21) in addition to \red{$3\sigma$} upper limits from the {\it Herschel Space Observatory} \citep{Pilbratt:10} across $100-500\,\mu$m \red{(see D21)}. However, we do present the ALMA 100-GHz flux density in Table~\ref{tab:opt_mir} as it plays a key role in constraining the photometric redshifts (see \S\ref{sec:fitting}). We note that the ALMA image presented in Fig.~\ref{fig:F105W} is the most sensitive produced by giving the data natural weighting, but our best estimate (D20) of the size of the radio source at 100\,GHz is $\le 0.7\,$arcsec from an image with Briggs (robust $=0$) weighting. The overall radio SED fitting in D20 and D21 did not suggest any variability in the radio and the $1.4\,$GHz flux densities from NVSS and FIRST are consistent within the uncertainties as are the two epochs of data from the VLA Sky Survey \citep{Lacy:20}. Hence, we conclude that this radio source is unlikely to be variable or to be beamed. 

\subsection{\red{Photometry}}
\label{sec:photanal}
We tied all \red{14 optical-to-mid IR ($0.48-22\,\mu$m)} images to the {\it Gaia} DR2 reference frame \red{using a catalogue obtained from the {\it Gaia} archive\footnote{\red{\url{https://gea.esac.esa.int/archive}}}}. 
\red{Catalogues derived from {\tt Source Extractor} were cross-matched ($\le 1\,$arcsec) to the {\it Gaia} catalogue and the derived mean right ascension and declination offsets (all $<0.2\,$arcsec) were added to the FITS sky position reference values of each image (i.e. CRVAL1 and CRVAL2 respectively). Note the offsets were less than $0.05\,$arcsec for {\it HST} and {\it HSC} so were not applied. As \gh~was not detected in the {\it GALEX} images we did not shift them. For the VIKING and {\it WISE} images we used the mean offsets of the four or five bands in each set, as these data have already been put onto an internally self-consistent reference frame.}

\red{Using the $0.7\,$arcsec radius apertures defined by the positions in Table~\ref{tab:pos}, we measured the flux densities, and uncertainties, in all bands\footnote{\red{Note that using {\tt Source Extractor} required all images, including the {\it HST} ones, to be on the same pixel grid, which we achieved by running {\tt swarp}, but only reprojecting the images down to the pixel size of the {\it HST} images ($0.12925\times 0.12925\,$arcsec).}} for the host of \gh~and the companion galaxy. The fluxes and uncertainties were directly converted from ADU to $\mu$Jy using the zero points and we applied an aperture correction to account for flux outside of the $0.7\,$arcsec aperture. The zero points and aperture corrections are reported in Table~\ref{tab:opt_mir} along with point-spread-functions (PSFs, $\theta_{\rm FWHM}$), the flux densities measured and their uncertainties.  To determine the aperture corrections for WFC3, we took the inverse of the nearest set of values of the fractional flux enclosed in Table 7.7 of the WFC3 instrument handbook\footnote{\url{https://www.stsci.edu/itt/review/ihb_cy17/WFC3/c07_ir7.html}} and interpolated them linearly using the $0.7\,$arcsec radius and central wavelength of each filter. For the HSC, VIKING and HAWKI data we empirically measured the curve of growth for nearby non-saturated stars to determine the aperture corrections.} 

\red{The {\tt Source Extractor} uncertainties are derived from the local RMS multiplied by the square root of the aperture area (in units of pixels). The resampling by {\tt swarp} affects the RMS through the algorithms used in the regridding process. To ensure we have accurate uncertainties, we have to apply a correction factor, $C$, to the uncertainties from the regridded images. On the assumption that the uncertainties should be the same for the original and regridded images,  one can show that $C=\frac{\sigma_{\rm o}}{\sigma_{\rm r}}\times f$ where $\sigma_{\rm o}$ is the pixel rms in the original image, $\sigma_{\rm r}$ is the rms in the resampled image, and $f$ is the scale factor by which the one-dimensional length of the square pixels has decreased by, i.e. $\theta_{\rm o}/0.12825$ where the $\theta_{\rm o}$ is the one-dimensional pixel length before regridding. We add an additional $10\%$ uncertainty, in quadrature, to all flux density measurements to account for the uncertainty in the absolute flux density scale and in the aperture corrections.}

\red{In the case of {\it GALEX} and {\it WISE}, we provide only $3\,\sigma$ upper limits (where $\sigma$ is the RMS noise) as \gh~ and the companion are clearly not detected and would be blended if they were (and in the case of {\it GALEX}, we did not correct to the {\it Gaia} reference frame). To determine the RMS for {\it WISE} we used the RMS estimated by {\tt Source Extractor} on the {\em original} image (i.e. before regridding, to avoid adding any correlated noise). These values were converted to AB magnitudes and then to $\mu$Jy. Determining the limit to the {\it GALEX} sensitivity was more complex than the other bands due to the low photon statistics. Indeed, the majority of pixels in the original FUV images contain no photons. Hence, we used {\tt Profound} \citep{Robotham:18} with specially tuned parameters to determine the RMS over an area with enough statistics and using mean not median statistics.
}

\section{SED fitting and Photometric Redshifts}
\label{sec:fitting}

In addition to the spectroscopy and template fitting from {\tt grizli}, we used 
three broad-band SED fitting methods to estimate the redshifts of \gh and the companion. These are (a) a bespoke UV-to-radio SED fitting method including the \red{measurements} and the upper limits, (b) {\tt EAZY} on the optical-to-near-IR detections, and (c) {\tt RAiSERed} on the radio photometry and imaging. 

\subsection{UV-to-radio SED fitting}
\label{sec:fittingfull}

Given the broad UV-to-radio wavelength coverage of \gh\, and building on our analysis in D21, we have developed a bespoke tool to simultaneously fit galaxy templates plus an analytical function to the SED. In this case the analytical function is a triple power-law (TPL) to fit the synchrotron emission dominating the radio SED at $0.07-100\,$GHz, whereas the galaxy templates cover the UV to far-IR. Note the template and power-law can both contribute to the 100\,GHz datum which motivates our approach of combining a template with an analytical model. For each class of galaxy template we determine the best fit of the combined template plus a TPL at 100 linearly-spaced redshifts across $0<z<10$.

\begin{figure*}[th!]
\begin{center}
\includegraphics[trim={1cm 0.35cm 2cm 2.4cm},clip, width=40pc, height=12.1pc]{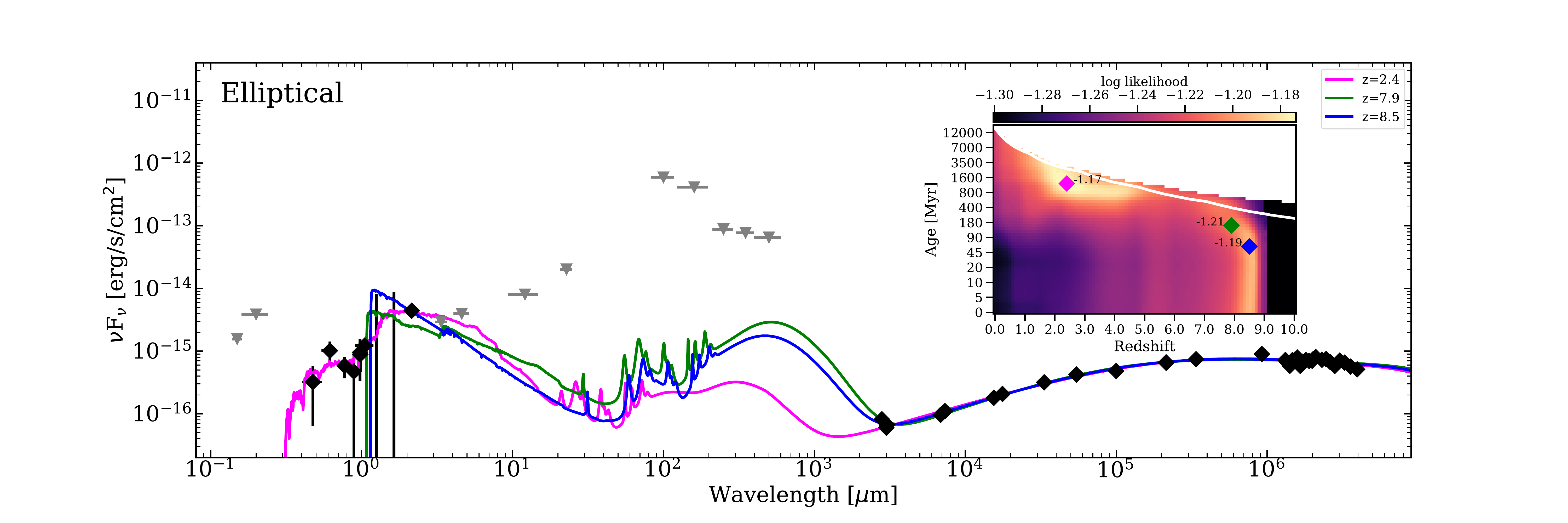}
\includegraphics[trim={1cm 0.35cm 2cm 2.4cm},clip, width=40pc, height=12.1pc]{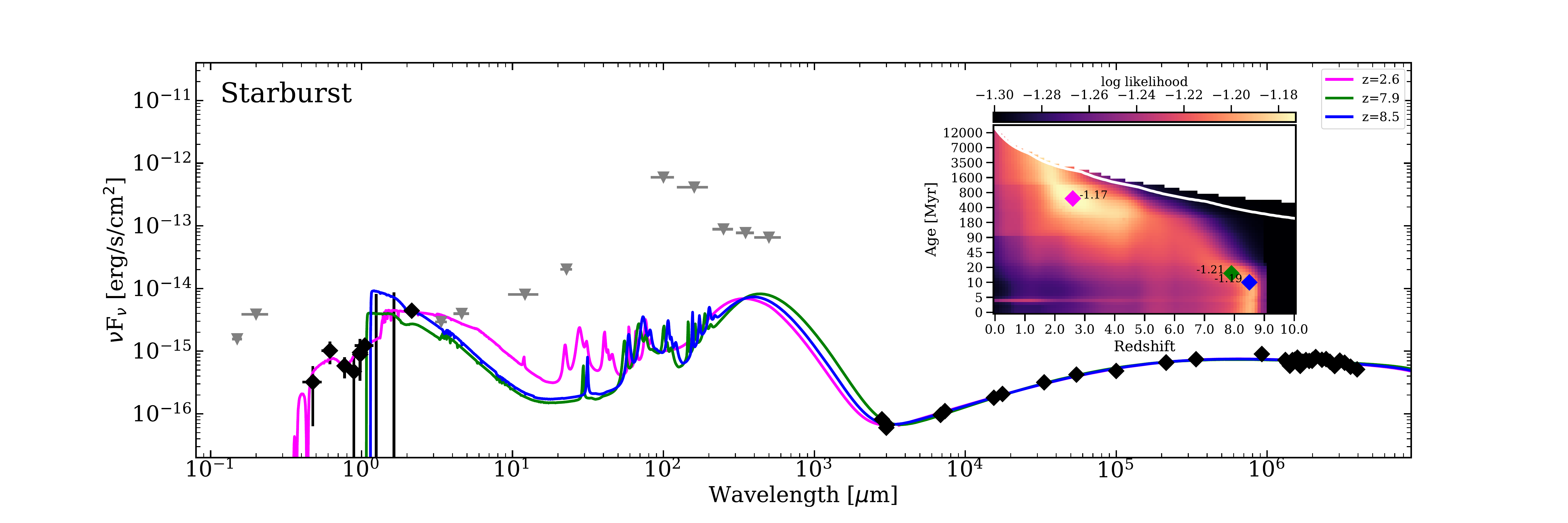}
\includegraphics[trim={1cm 0.35cm 2cm 2.4cm},clip, width=40pc, height=12.1pc]{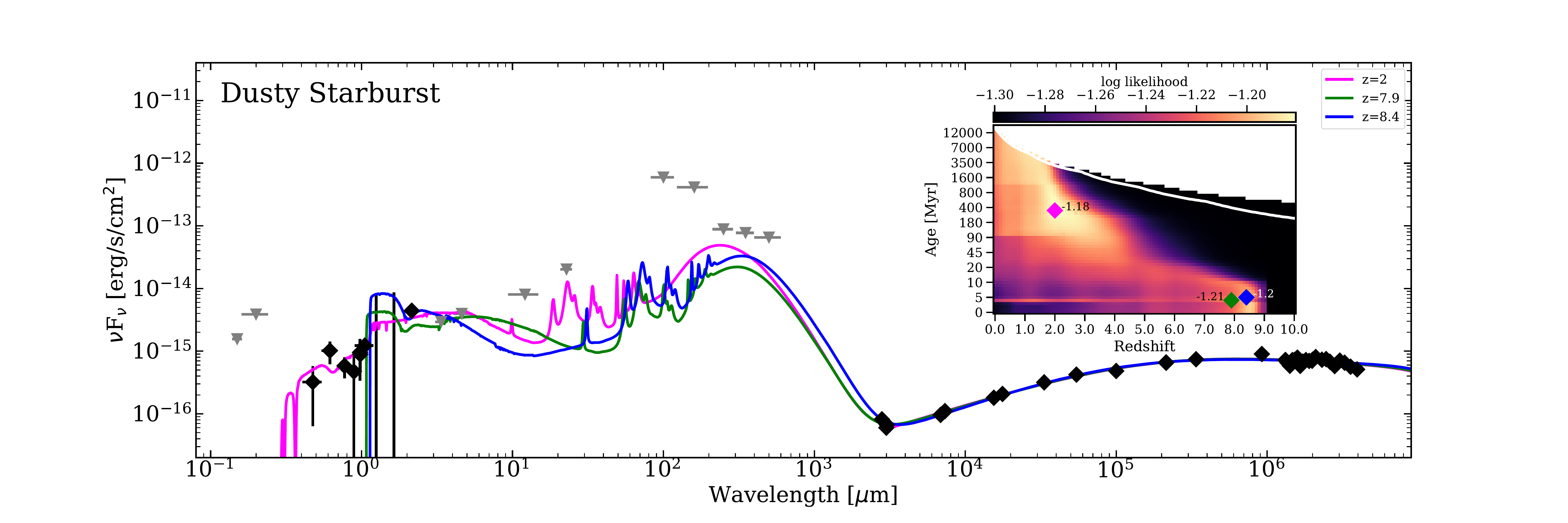}
\includegraphics[trim={1cm 0.35cm 2cm 2.4cm},clip, width=40pc, height=12.1pc]{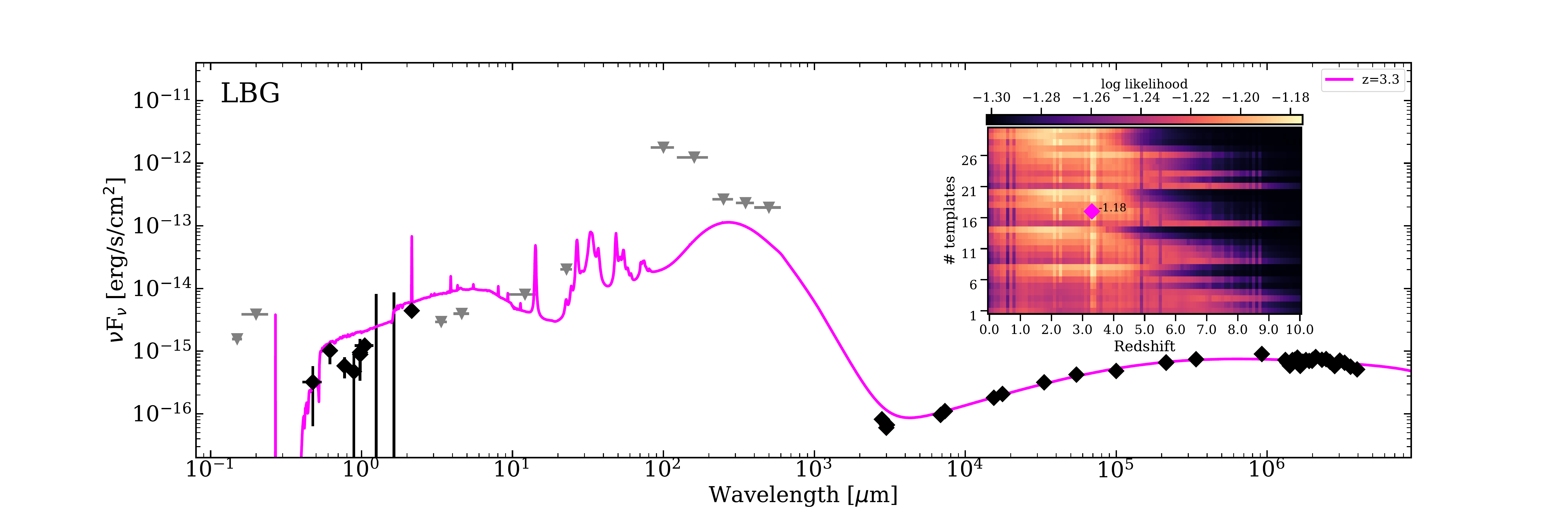}
\vskip -0.5cm
\caption{Observed UV-to-radio SED of \gh~including \red{measurements (black diamonds with $1\sigma$ errorbars) and $3\sigma$} upper limits (triangles) overlaid with a selection of best-fitting models. Each panel corresponds to a different galaxy template class and the radio data is simultaneously fit with a synchrotron triple-power-law model (see text for full details). The insets show the distribution of likelihood values for each redshift/age combination (redshift/template combination for the LBG templates). Omitted are solutions which exceed the age of the Universe (i.e the top right of the insets \red{for the evolving PEGAS\'E tempaltes}). The curved white line indicates when the first galaxies are thought to have formed \citep[e.g.][]{Laporte:21}, around 250\,Myr after the Big Bang. The symbols in the insert\red{, with their log likelihoods labelled} correspond to local \red{maxima} from which the best-fit templates are taken (plotted in the same colour). This modelling suggests an old, low-redshift ($z\sim2$\red{, in magenta}), or a high-redshift ($z\sim 7.9-8.5$\red{, in green and blue}) solution is most preferred.}
\label{fig:0917fitP}
\end{center}
\end{figure*}

We employ the following templates from {\tt P\'EGASE} \citep[v3;][as used in D21]{Fioc:19} with the addition of some Lyman-break galaxy (LBG) ones:

\begin{enumerate}
    \item An elliptical galaxy template.
    \item A starburst template with internal dust extinction and assuming a slab geometry.
    \item \red{A dusty starburst template}, i.e. the same starburst template but with $\alpha_{\rm slab}$=10, corresponding to a column density ten times denser\footnote{See \S\,I.6.c of the {\tt P\'EGASE.3} manual: \url{http://www2.iap.fr/users/fioc/Pegase/Pegase.3/}}.
    \item The 30 empirical LBG templates from \cite{Alvarez-Marquez:19} which cover a range of UV luminosities, UV continuum slopes and stellar mass. 
\end{enumerate}

All these templates cover from below the Lyman limit to the far-IR ($0.01<\lambda_{\rm rest}<10\,000\,\mu$m). For the {\tt P\'EGASE} templates we take 60 logarithmically-spaced ages across $0-3,500\,$Myr, although exclude redshift/age combinations which exceed the age of the Universe. Hence, for each galaxy template (plus TPL) we calculate the highest likelihood, and hence best-fit, for all 6,000 redshift/age combinations. For the 30 LBG templates, the parameter space explored covers 3,000 redshift/template combinations. 

For the TPL, we use Equation 4 from D20. The initial parameters to the triple power-law were those reported in Table 3 of D21, but with the low-frequency spectral break and low-frequency slope fixed. Importantly this allows the parameters in the triple power-law which effect the 100\,GHz \red{datum} to vary as required \red{in order} to accommodate the galaxy template. Hence, for each of the 9,000 fits, the parameters varied are the template normalisation plus four out of six TPL parameters. 

When fitting the data we use the same likelihood calculations as used in  the {\tt MrMoose}\footnote{\url{https://github.com/gdrouart/MrMoose}} SED fitting code \citep{Drouart:18}. This formalism has the advantage of being able to use 
\red{both flux measurements, even at low or negative SNR, and flux upper limits, as we use in the lower resolution bands. The full details of this approach are explained in the appendix of \citet{sawicki:12}, but, briefly, the minimised $\chi^2$ is the sum of the traditional $\chi^2$ statistic for the flux measurements and an additional term for the upper limits. This additional term accounts for the probability that the observation in a given band is drawn from a given model. As such, it is an integral from minus infinity to the user defined upper limit. Here we choose $3\sigma$ as the upper limit although the fitting is not very sensitive to this limit. Effectively all the upper limits} contribute to the logarithm of the likelihood (analogous to $\chi^2$ if using detections only), \red{i.e. the contribution to the likelihood} progressively penalises models that lie \red{above the $1\sigma$ RMS noise and up to the predefined limit (here $3\sigma$)} and favours those \red{models that lie} below the limits \citep[see][for more details]{Drouart:18}. 

\begin{figure*}[th!]
\begin{center}
\includegraphics[trim={1cm 0.35cm 2cm 2.4cm},clip, width=40pc, height=12pc]{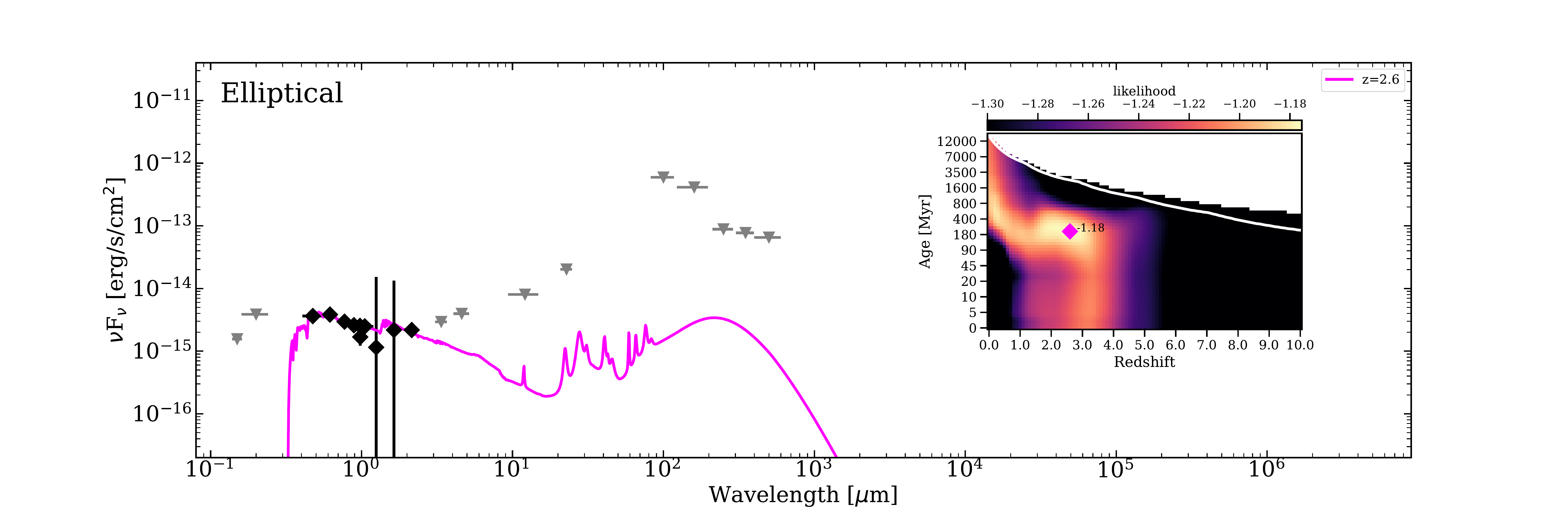}
\includegraphics[trim={1cm 0.35cm 2cm 2.4cm},clip, width=40pc, height=12pc]{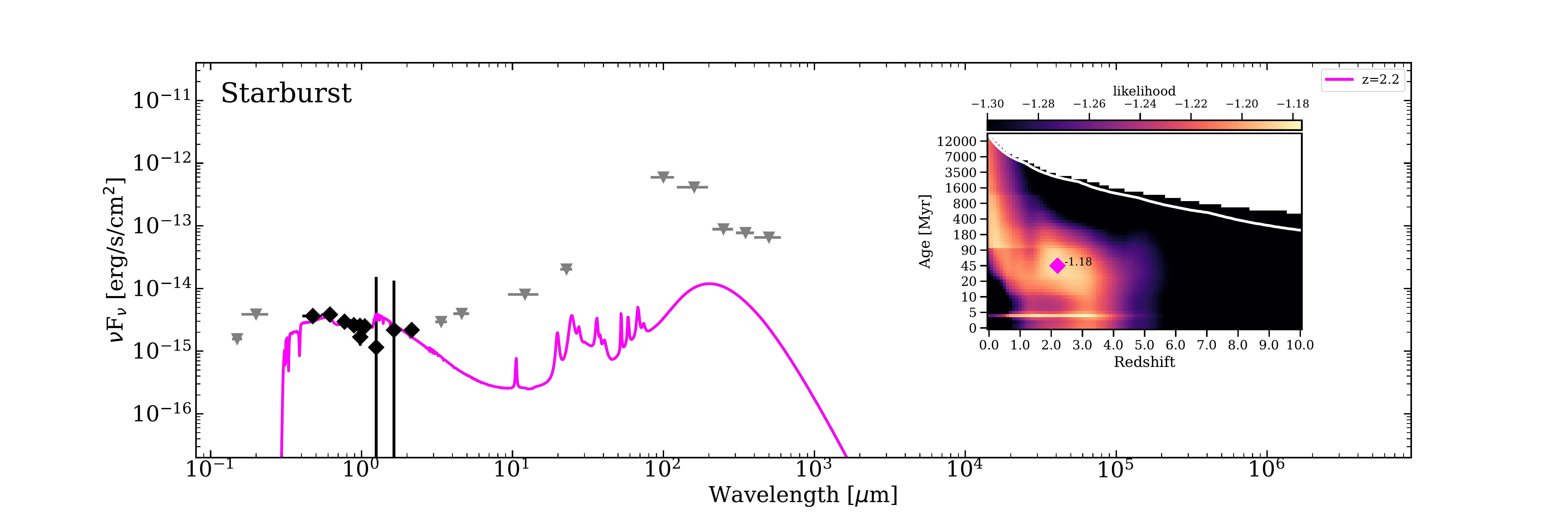}
\includegraphics[trim={1cm 0.35cm 2cm 2.4cm},clip, width=40pc, height=12pc]{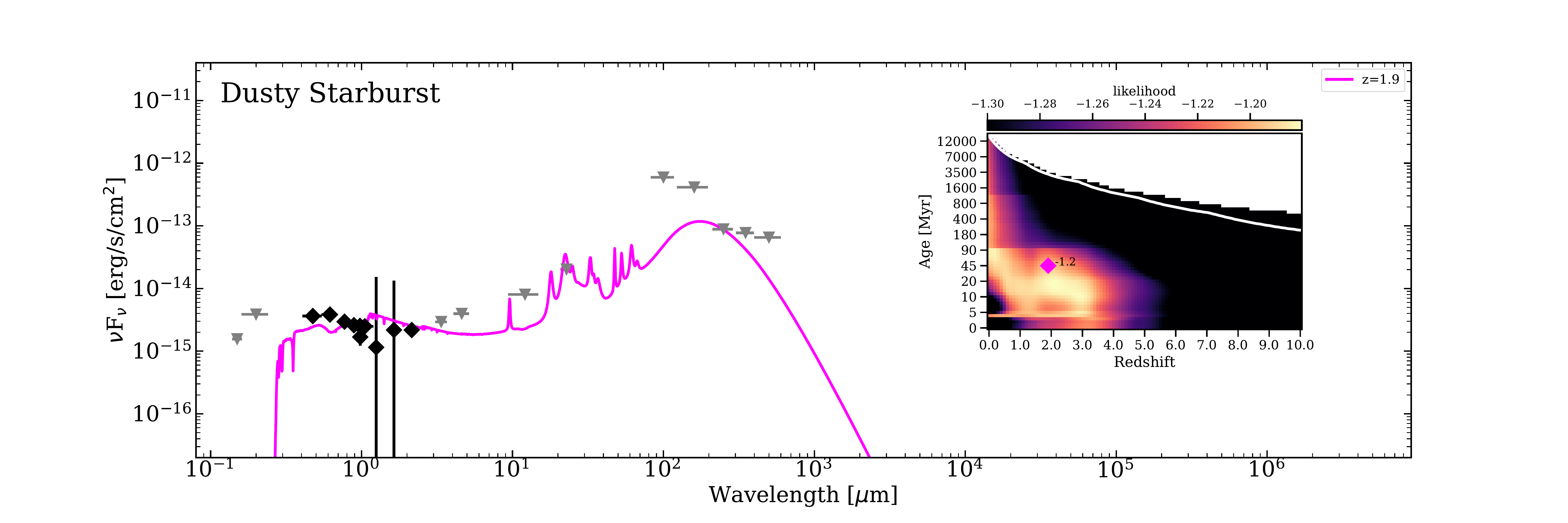}
\includegraphics[trim={1cm 0.35cm 2cm 2.4cm},clip, width=40pc, height=12pc]{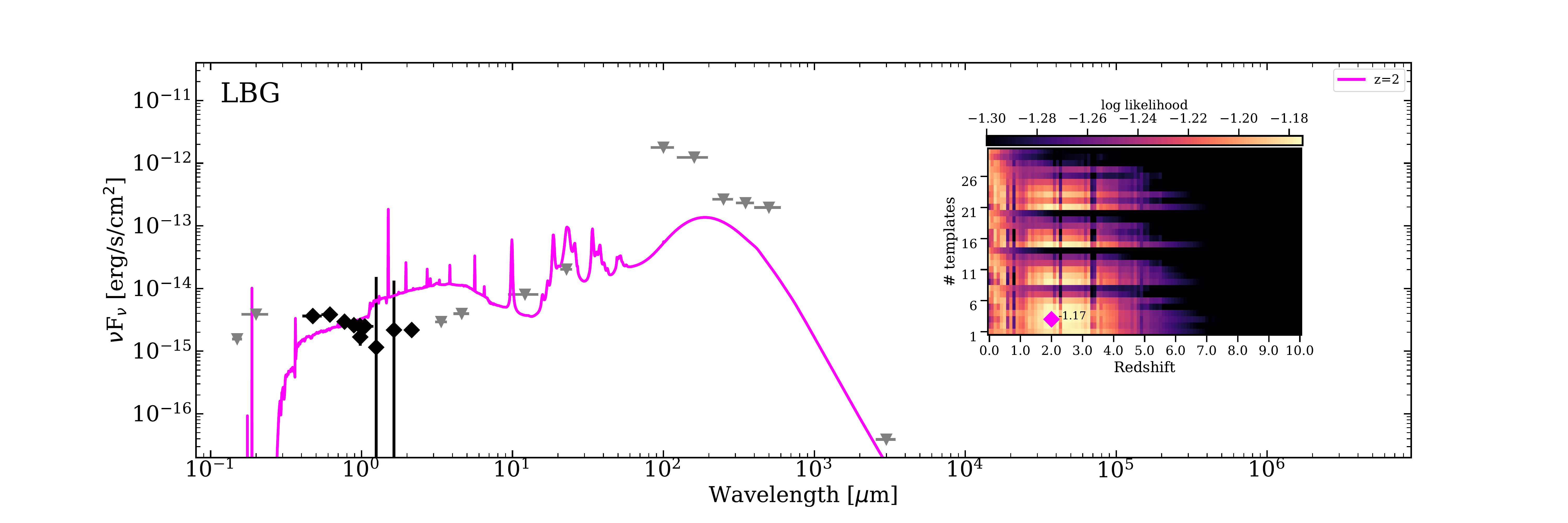}
\vskip -0.5cm
\caption{As Fig.~\ref{fig:0917fitP}, but for the companion galaxy. As this source is radio-quiet the templates are fit without a synchrotron model. Broadly speaking, the best solution\red{s are at $z< 3$ (magenta)}.}
\label{fig:compfit}
\end{center}
\end{figure*}

We present a selection of the best-fitting SEDs to \gh\ for each template class in Fig.~\ref{fig:0917fitP}. We fit the models in flux density space, but the resultant SEDs are plotted in power, i.e. $\nu\,F_\nu$, to better illustrate the SED over almost eight decades in wavelength. The inserts show the variation of likelihood across the age/redshift parameter space (template/redshift parameter space for \red{the LBG templates}). The SEDs plotted are from representative local \red{maxima} in the likelihood distribution. We also use the same code, minus the analytical function \red{for the synchrotron emission}, to estimate the redshift of the radio-quiet companion galaxy (see Fig.~\ref{fig:compfit}).

\subsection{Optical-to-near-IR SED fitting with {\tt EAZY}}
\label{sec:eazy}
Using the optical-to-near-IR detections we can implement the more traditional {\tt EAZY} (version 2015-05-08) photometric redshift estimator \citep{Brammer:08}. \red{Unlike the bespoke fitting above, {\tt EAZY} cannot use the lower resolution data where we only have upper limits. Hence, we ran} {\tt EAZY} on the host galaxy and the companion using the data in Table~\ref{tab:opt_mir} \red{bar the lower-resolution UV and mid-IR data}. We used the default templates \red{provided and the} redshift priors (linked to the $K_{\rm s}$-band magnitude in this case) \red{although they make little difference to the best-fit redshift.} We apply the intergalactic medium (IGM) absorption option as we allow the estimated redshift to cover $0.01<z<10$. \red{We report the marginalised redshift estimates with the prior, although the prior only has a weak effect on the final result.}

\subsection{Radio continuum redshift fitting with {\tt RAiSERed}}
\label{sec:raisered}

\citet{turner:20} proposed a dynamical-model based method, called {\tt RAiSERed}, for determining the redshift of a radio galaxy using exclusively radio-frequency imaging and photometry. In this code, the breadth of radio galaxies permitted by the physics of the Radio AGN in Semi-analytic Environments \citep[{\tt RAiSE};][]{Turner:15,Turner:18a} dynamical model are compared to the properties of a given radio galaxy being investigated, yielding a probability density function for its redshift. Specifically, one can compare the lobe flux density, angular size and broad-band spectral shape of the dynamical model and observed source. 
\red{The local environment of the radio galaxy is modelled as a prior probability density function based on the local cluster mass function \citep{girardi:00}, modified for higher redshifts using semi-analytic galaxy evolution models \citep[SAGE,][]{croton:16,Raouf:17}. Hence with an increasingly dense environment at higher redshifts, degeneracies with redshift in the model are avoided.}

We apply this method to \gh, taking the flux density of each lobe at 151\,MHz as $0.233\pm0.014$\,Jy (half the total flux density presented in D20) whilst the angular size of each lobe is $<0.35$\,arcsec\footnote{The source (both lobes) is unresolved, including at millimetre (ALMA) wavelengths which has a $0.7\,$arcsec resolution.}. The \red{unknown} lobe axis ratio is assumed to be consistent with that of Cygnus A \citep{Turner:19} as a representative powerful radio galaxy.

The broad-band spectral shape across the $0.07-20\,$GHz radio data \red{may be} described \red{by} two different models:
\begin{enumerate}
    \item lobe-dominated
    \item young-jet \red{plus} aged-lobe
\end{enumerate}

The lobe-dominated model assumes a population of syn\-chrotron-emitting electrons shock-accelerated uniformly in time since the commencement of jet activity, and which presently have \red{an} approximately constant magnetic field strength. The resulting spectrum is well described by the continuous injection \red{of relativistic electrons} and characterised by a power-law at low frequencies that steepens beyond an (optically-thin) break frequency by $\Delta \alpha = -0.5$\footnote{\red{The radio spectral index, $\alpha$, is defined by $S_\nu\propto\nu^\alpha$.}} for active sources. The initial power-law was fit in D20 \red{and D21} as $S_\nu \propto \nu^{-0.8\pm0.1}$ with the optically-thin spectral break at $1.6_{-0.6}^{+0.9}$\,GHz. However, \red{at any redshift, the lobe-dominated} model poorly explains the steep spectral index within the MWA band (100-150\,MHz) of $S \propto \nu^{-1.0\pm0.1}$, albeit is consistent with the low-frequency spectral index of the continuous injection model at the 2$\sigma$ level.

\red{For high-redshift radio sources, the much stronger cosmic microwave background radiation will rapidly deplete the energy of the synchrotron electron population in the lobe through inverse-Compton scattering. Hence, the jet contribution to the integrated luminosity will be important. Therefore, we also fit a young-jet plus aged-lobe model.} Electrons in jets are continually accelerated at shocks, both internal (such as recollimation shocks) and termination shocks (i.e. hotspots), resulting in power-law spectra characteristic of the particle energy spectrum at injection \citep[e.g.][]{Bicknell:18}. These spectra age rapidly on cessation of particle acceleration, on timescales $<1$~Myr at GHz radio frequencies, due to the high milli-Gauss magnetic fields in the jet. Many powerful radio galaxies are intermittent \citep[e.g.][]{Bruni:19}, and we therefore use only the low-frequency ($<1.4\,$GHz) radio data to fit the jet injection spectrum. \red{The higher-frequency observations are therefore modelled assuming either a continuous injection or a \citet[][JP]{Jaffe:73} spectrum.}

Hence, the second model considers the integrated luminosity for such sources using a two-component spectrum: \red{a young jet plus aged lobe}. The young jet is assumed to have only recently accelerated synchrotron-emitting electrons, \red{but is described by a JP continuous injection spectrum due to the extreme energy losses present at high redshift. These losses are due to a combination of the high magnetic field strength and strong inverse-Compton field which varies with redshift}.
Meanwhile, the aged lobe is described by a continuous injection model but with a break frequency well below the lowest observing frequency\red{, i.e. it has a power-law break} $\Delta\alpha=-0.5$ steeper than the jet. This two-component spectrum is fitted assuming arbitrary flux density scaling for the two components and the same initial spectral index upon shock-acceleration of synchrotron-emitting electrons. We constrain the initial power-law spectrum of both components as $S_\nu \propto \nu^{-0.8\pm0.1}$ and the break frequency of the aged-lobe component as $<100$\,MHz\red{; this model fits a spectral index at 100\,MHz of $S_\nu \propto \nu^{-1.0\pm0.1}$, consistent with the value fitted across the MWA band. The results of this fit are shown in Fig.~\ref{fig:raisered}. } 

\red{We further consider the possibility that \gh~ is a remnant radio source. The {\tt RAiSERed} model fitting currently does not allow for an inactive source, so this approach cannot constrain the redshift, but only determine if this is a feasible model for the radio spectrum. We start by fitting the spectrum with the \citet[][]{komissarov:94} model. This model yields a comparable power-law spectrum at low frequencies, but which steepens earlier than the lobe-dominated model for an active source such that it is $S \propto \nu^{-0.9\pm0.1}$ at 100\,MHz. The off-time is fitted as $13\pm3$\% of the source total age under the assumption it is a remnant source. We discuss why this model is unlikely astrophysically in \S\ref{sec:res_raisered}.}

\begin{figure*}[t!]
\begin{center}
\includegraphics[width=20pc, height=15pc]{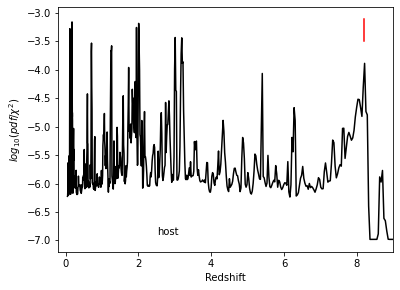}
\includegraphics[width=20pc, height=15pc]{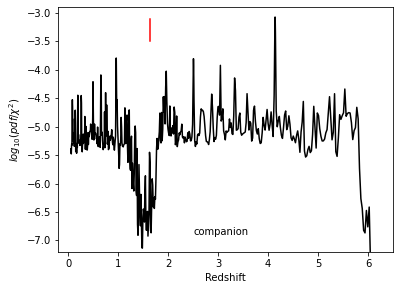}
\vskip -0.5cm
\caption{{\tt Grizli} redshift probability distribution function divided by the $\chi^2$, up to $z=9$ for the \gh~and up to $z=7$ the companion. Both distributions have many maxima occurring when one line or another is matched to the lines detected in the host and companion. In both cases we highlight (by a red vertical line) the solution which is most consistent with the SED fitting methods and is from a strong line. For example, we discount the $z=4.14$ maxima for the companion as this would mean we have detected the [CIII]/$1909\,$\AA~ line which is very weak. Hence, our best solution for the host galaxy is $z=8.21$ \red{(Lyman-$\alpha$)} and for the companion $z=1.63$ \red{(C[II]])}.}
\label{fig:grizlizpdf}
\end{center}
\end{figure*}

\section{Results and Discussion}
\label{sec:results}
\subsection{Imaging and Photometry}

The host of \gh~is detected and resolved in our F105W image (Fig.~\ref{fig:F105W}) and is detected less strongly in the F098M band (Table~\ref{tab:opt_mir}). However, even with the priorised fitting the host of \gh~remains undetected in the HSC (shortward of the {\it HST} data) and VIKING bands. The HSC and HAWKI flux densities reported in Table~\ref{tab:opt_mir} for the companion are consistent with those from D21 to within $10\%$. For the host, the $K_{\rm s}$-band flux density is found to be $16\%$ lower (including the aperture correction on the flux density reported here) than that found by D21 which is most likely due to the matched aperture photometry used to measure it (although it is consistent within $2\sigma$). 

The host is mainly compact at $0.25''$ across, but with a faint extension to the west towards the `companion' (believed to be at a different redshift; see \S~\ref{sec:disc}). The ALMA 100-GHz continuum contours overlaid in Fig.~\ref{fig:F105W} clearly identify the eastern source as the host (as determined in our original analysis; see D20 and D21). This emission is potentially aligned with the faint extension to the west, but the ALMA resolution makes this inconclusive. The UV-to-radio SED fitting (\S\ref{sec:resfullsed}) suggest that there could be both synchrotron and cold dust contributing to this emission. 

The companion source is more extended ($\sim 0.3''\times 0.4''$) and brighter in the F105W band. It is also detected in all five HSC bands including $g$-band, ruling out that this source is above $z\sim 4.2$ (using the Lyman limit). The flux densities measured here are consistent with the catalogued values from HSC to within $8\%$. 

\subsection{Spectral Properties and {\tt grizli} Template Fitting}

\subsubsection{\gh}

The {\tt hstaxe} and {\tt grizli} spectra (Fig.~\ref{fig:0917spec}) both show a faint, broad emission feature at $1.12\,\mu$m and no discernible continuum. \red{The {\tt grizli} SED fitting routine estimates a SNR of $\sim 3$ for this feature (see Table~\ref{tab:line}).} The {\tt hstaxe} extraction also shows a brighter, narrow feature at $1.15\,\mu$m, but this is not seen in the {\tt grizli} reduction. We do see it in both orientations of the {\tt hstaxe} reduction but it is likely the combined effect of cosmic rays and/or hot pixels. Furthermore, this feature occurs on the part of the G102 transmission where the throughput is decreasing very rapidly towards longer wavelengths hence we do not consider it a reliable feature. 

As well as producing spectra, {\tt grizli} fits templates at a range of redshifts in order to identify any features and confirm the redshift. One output from this analysis is the PDF. We show in the left panel of Fig.~\ref{fig:grizlizpdf} the PDF divided by $\chi^2$ for the host. Numerous maxima occur at different redshifts, but we highlight the one we believe to be most likely at $z=8.21$ (if the $1.12-\mu$m feature is Lyman-$\alpha$) considering the strength of the identified line and the constraints from the SED fitting (\S\ref{sec:resfitting}). Curiously, the NV$\lambda\lambda 1238,1242$ line is found in absorption but we do not give this much credence due to the low SNR at that wavelength. The other most likely solution is $z=2.01$ (if the line is the [OII]$\lambda 3727$ doublet). 

\subsubsection{Companion}

The {\tt hstaxe} and {\tt grizli} spectra (Fig.~\ref{fig:comspec}) both show a faint, broad emission feature at $0.98\,\mu$m and no discernible continuum. The PDF over $\chi^2$ from the {\tt grizli} template fitting (right panel of Fig.~\ref{fig:grizlizpdf}) also shows numerous solutions. The most probable solution from a naive examination of the {\tt grizli} PDF is $z=4.14$ (which depends on the $0.98$-$\mu$m feature being the C[III] line at $\lambda_{\rm rest}=0.1909\,\mu$m). Given the relative weakness of this line we think this solution to be unlikely and discuss the other possibilities in \S\ref{sec:disc}. 

\subsection{SED Fitting}
\label{sec:resfitting}

We used three methods to estimate the redshift from the broad-band photometry. The results from {\tt EAZY} and {\tt RAiSERed} are best summarised numerically in Table~\ref{tab:z} where the uncertainties or limits (where available) are given as the $68^{\rm th}$-percentile.

\subsubsection{UV to Radio SED Fitting}
\label{sec:resfullsed}

For the host galaxy, the results of this fitting (Fig.~\ref{fig:0917fitP}) for the different \red{PEGAS\'E} templates \red{(top three panels} are broadly consistent.  The insets in these figures show the most preferred redshift solutions and there are two consistent \red{maxima} at $z\sim2$ and  $z\sim 8$. The low-redshift solution is consistent with an older galaxy \red{(with an age of} several \red{hundred} Myr. The high-redshift solution is quite broad, $7.5<z<8.5$, and generally prefers a \red{young} age of \red{$\sim 10\,$}Myr. Overlaid on the data in Fig.~\ref{fig:0917fitP} are the best-fit templates at \red{$z=2.4, 7.9$ and $8.5$} which \red{are} all comfortably consistent with the photometry. These results \red{also agree well} with the redshift constraints in the optical-to-IR SED fitting of D21 which found a low ($z\sim2$) and high ($z>7$) redshift solution. 

\red{As seen by the colour-bar for the parameter space `heatmap' inserts, while there are maxima as highlighted above, the overall variation in likelihood is not large. Hence, overall no particular redshift is strongly preferred by this method.
}

The LBG templates (Fig.~\ref{fig:0917fitP}, bottom panel) are broadly consistent with the lower-redshift solution \red{$z<3$. However, the best fit template does not appear to be good representation of the data.}

The synchrotron component is not significantly affected by the fitting of the different template classes. Only the high-frequency slope and high-frequency spectral break change with differing template contribution to the 100-GHz datum. However, the best-fitting results suggest that this far-IR datum has a significant cold dust contribution. The frequency at the \red{low-frequency} turn-over (in $S_\nu$ space) and low-frequency slope are fixed from a fit to just the $0.07-100\,$GHz data. The observed frequency of the turn-over is constrained to be $<50\,$MHz with a 90\% confidence although it appears higher in the figures as we plot in $\nu\,S_\nu$ \red{(rather than $S_\nu$)}.

For the companion galaxy the results of this fitting (Fig.~\ref{fig:compfit}) for the different templates are also broadly consistent between the top three panels. These solutions generally favour a `low'-redshift ($z<3$) result compared to the potential high-redshift solution for the host galaxy. In these panels, there is a broad distribution of high likelihoods across $0<z<3.5$. A local \red{maxima is} highlighted in Fig.~\ref{fig:compfit} with the corresponding template overlaid on the data. Likewise with the LBG templates there is broad region of high likelihood. These results on their own do not further constrain the redshift of the companion.

\subsubsection{\tt EAZY}
\label{sec:res_eazy}

We are able to run {\tt EAZY} on both the host and companion using only the optical to near-IR \red{measurements}. The results are reported in Table~\ref{tab:z}. For the companion, {\tt EAZY} finds a low-redshift solution of $z\sim  2.6$, compatible with the {\tt EAZY} result from D21 within the uncertainties. For the host galaxy of \gh, {\tt EAZY} finds a result ($z\sim 2.4$) albeit from \red{a higher fraction of low SNR measurements}. We could not use {\tt EAZY} \red{on the host} in D21 with only one detection. 

\begin{table}[t!]
\begin{threeparttable}
\caption{Constraints on the redshift of the host and companion using the {\tt EAZY} and {\tt RAiSERed} (which is only possible for the host). We also show for reference the results of {\tt EAZY} applied to the companion in D21. The uncertainties are the $68^{\rm th}$-percentile confidence intervals from the resulting redshift distributions.} \label{tab:z}
\tabcolsep=8pt

\begin{tabular}{ l c c } \toprule
Photoz & \multicolumn{2}{c}{Redshift Estimate} \\
Method & Host & Companion \\ \midrule
{\tt EAZY}\,(D21) & --- & $2.2^{+0.3}_{-0.6}$ \\
{\tt EAZY}        & $2.42^{+0.75}_{-0.74}$ & $2.56^{+0.51}_{-0.54}$ \\
{\tt RAiSERed} & $>6.5$\tnote{a} & n/a \\ \bottomrule
\end{tabular}
\begin{tablenotes}
\item[a] From the  young-jet \red{plus} aged-lobe model.
\end{tablenotes}
\end{threeparttable}
\end{table}

\begin{figure*}[t!]
\begin{center}
\vskip -0.5cm
\includegraphics[width=20pc, height=15pc]{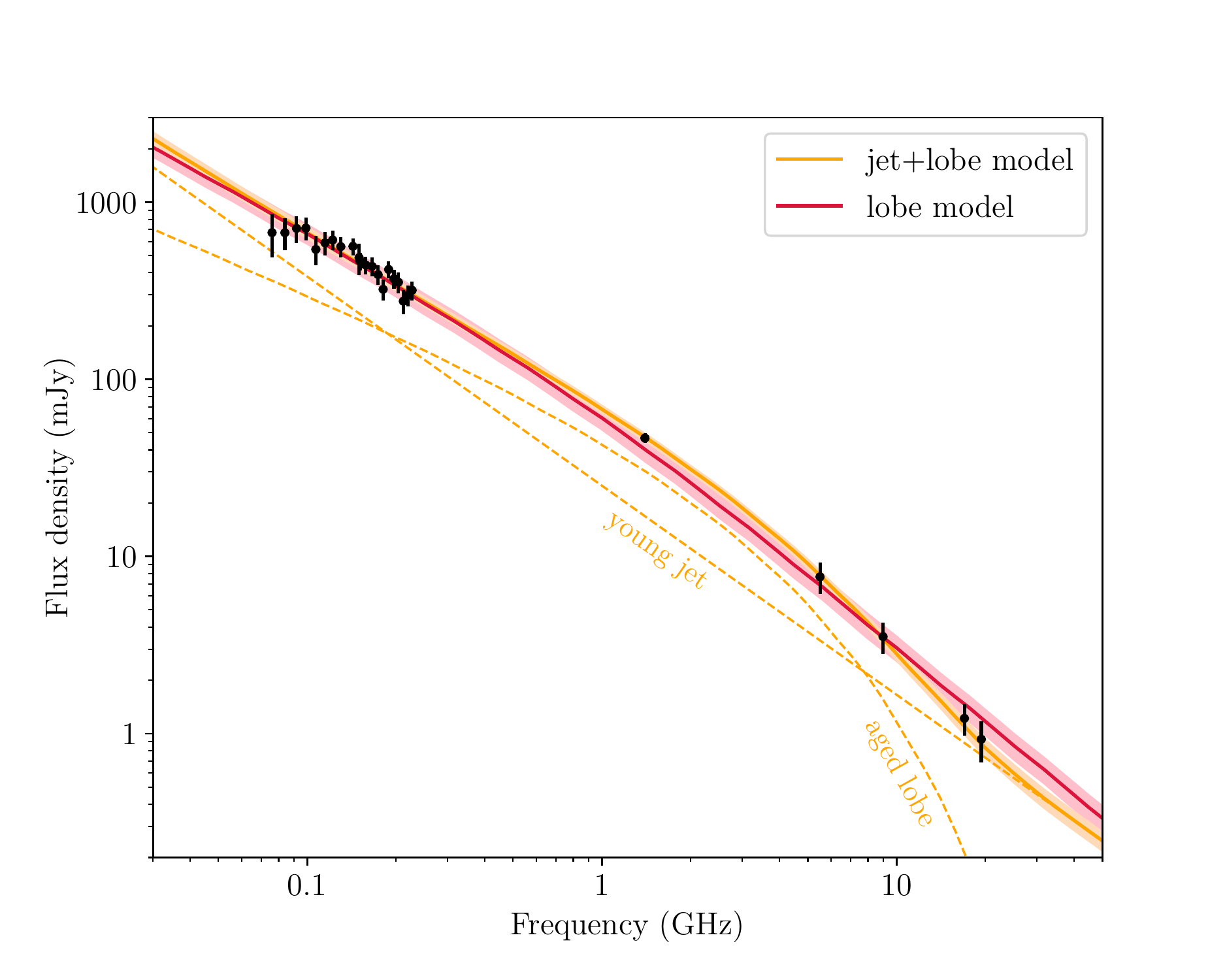}
\includegraphics[width=20pc, height=15pc]{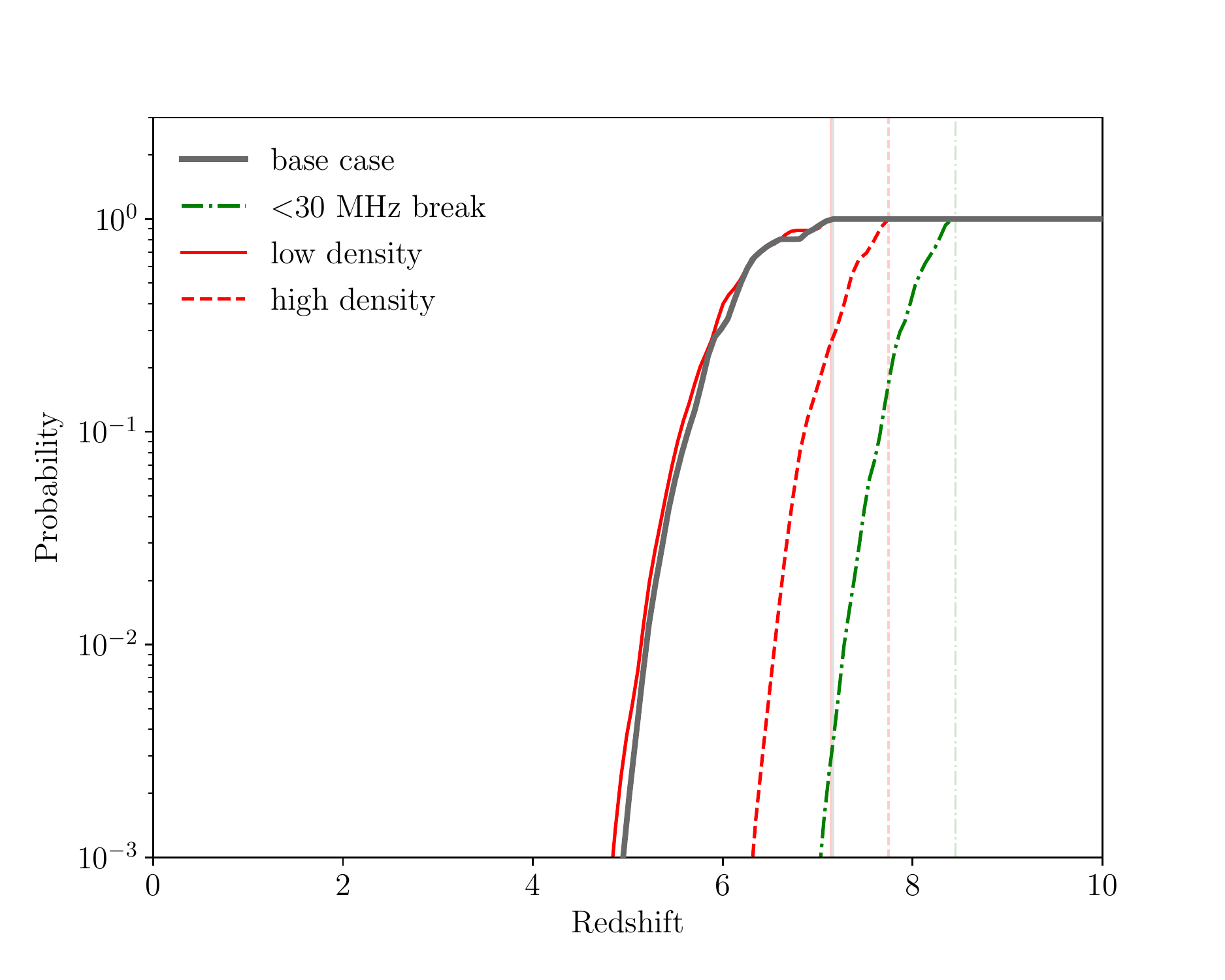}
\vskip -0.5cm
\caption{\red{Results of the {\tt RAiSERed} radio SED  fitting. In the left panel we show the two best fit models: a lobe model at $z\sim 2.7$ and a young-jet plus aged-lobe model at $z\sim 7$. For the jet-plus-lobe model we show the two components separately, with the intermittent jet being the component dominant at higher frequencies. In the right panel we present the PDF for the redshift of the young-jet plus aged-lobe model with the smoothed distribution in black peaking for all redshifts greater than $\sim 7$. This fit should be considered an upper limit as we have assumed the jet component is symmetric when it may not in fact lie in the plane of the sky (see \S\ref{sec:res_raisered} for more details). The green dot-dashed line assumes a lower break frequency of 30\,MHz, whilst the two red lines increase/decrease the density in the ambient media assumed in {\tt RAiSERed} by a factor of two. Note we do not fit the low-frequency turn-over as it has no impact on the RAiSERed results. }}
\label{fig:raisered}
\end{center}
\end{figure*}

\subsubsection{\tt RAiSERed}
\label{sec:res_raisered}

\red{The {\tt RAiSERed} fitting} presents us with two different models with which to \red{fit} the radio emission of \gh~and hence estimate the redshift\red{: a lobe-dominated and a young-jet plus aged-lobe model. We present our results in Table~\ref{tab:raisered}, as well as the spectral fitting results for a remnant model (no redshift constraint is available as {\tt RAiSERed} assumes active sources).} The lobe-dominated model constrains its redshift to be $z = 2.74\pm0.24$. However, this model has difficulties explaining the steep spectrum of the SED before the low-frequency turn-over. It also does not allow for the contribution of a jet to the total radio emission as is common for compact sources. The young-jet plus aged-lobe model gives a lower limit on the redshift of \gh~ as $z > 6.5$ under the assumption of a two-component spectrum. In this case, we would get $\sim 50\%$ jet prominence at \red{$\sim 150\,\rm MHz$}, consistent with the high values expected from compact sources \citep{Hardcastle:98}. \red{Figure~\ref{fig:raisered} presents both the best-fit SEDs for the young-jet plus aged-lobe and lobe-dominated {\tt RAiSERed} models, as well as the PDF of redshift from the young jet plus aged-lobe model.}

\red{While the Akaike Information Criterion \citep[AIC,][where lower values imply a preferred fit]{akaike_new_1974} significantly prefers the young-jet plus aged-lobe over the lobe-dominated model ($\Delta$(AIC)$>20$), the remnant spectrum has the lowest AIC of the three models. However, following \citet[][their Equation 19]{Turner:19}, we estimate that the magnetic field strength of \gh~is $>$10.6\,nT assuming a spherical source of angular size 0.7\,arcsec at $z = 2.01$ \citep[we assume the equipartition factor derived by][for 3C sources]{Turner:18a}; this increases to $>$18.5\,nT assuming the axis ratio of Cygnus A. For these magnetic field strengths, the optically-thin spectral break frequency of the remnant spectrum implies an active age of $<$0.51\,Myr and $<$0.23\,Myr respectively \citep[e.g. Equation 4 of][]{Turner:18b}. The jet is unlikely to have fully escaped its host galaxy on such timescales and therefore we would expect strong free-free absorption in our $<$1\,GHz observations \citep{Bicknell:18}.}

\begin{table*}[ht]
\begin{threeparttable}
\caption{\red{Results of the radio SED fitting with {\tt RAiSERed} and a remnant model. For each model we present the best-fit values for the low-frequency spectral index, ($\alpha_{\rm low}$ from the electron energy distribution, i.e. $N(E)\propto E^p$ and $p=2\times\alpha_{\rm low}-1$), the break-frequency  ($\nu_{\rm break}$), and, in the case of the remnant model, the fraction ($T$) the age where the jet has been turned off. We also give the resultant value of the Akaike Information Criteria (AIC) from the fit and the constraints on the redshift for the two {\tt RAiSERed} models.}} 
\label{tab:raisered}
\tabcolsep=6pt
\begin{tabular}{lccccc} \toprule
Model & $\alpha_{\rm low}$ & $\nu_{\rm break}$ & T($t_{\rm off}/\tau$) & AIC & Redshift \\
      &          & [GHz]  &  &  &  \\

\midrule
lobe-dominated & $0.88\pm0.07$ & $1.78\pm0.11$ & \ldots & 260.7 & $2.74\pm0.24$\\
young-jet plus aged-lobe & $0.693\pm0.0001$ & $<0.1$ & \ldots & 241.1 & $z>6.5$\\
remnant lobe$^a$ & $0.80\pm0.01$ & $1.82\pm0.12$ & $0.13\pm0.03$ & 234.5 & \ldots\\
\bottomrule
\end{tabular}

\begin{tablenotes}
\item[a] {\tt RAiSERed} is only applicable for active sources, so no redshift constraint is available.
\end{tablenotes}
\end{threeparttable}
\end{table*}

Another constraint from the radio emission comes from assuming that this radio galaxy is a peaked source lying on the peak rest-frame frequency versus physical size relation \citep{odea:97}. This relation states that as compact radio sources expand and grow, their peak frequency shifts to lower values. The fitting of the radio SED results in an observed-frame turn-over frequency of $<50\,$MHz at a $90\%$ confidence (consistent with D20 and D21). In Table~\ref{tab:line} we report the $1.4$-GHz radio luminosity, the rest-frame peak frequency upper limit and hence the inferred physical size lower limit (assuming the source lies on the peak/size relation) for both the potential redshifts. Note, the uncertainty on the peak frequency dominates the uncertainty in the peak/size relation. We know that at any redshift $>1.6$ our $0.7\,$arcsec radio size limit makes the projected size of this source $<6\,$kpc. The higher ($z\sim 8$) redshift solution is consistent with this limit, while the $z\sim 2$ solution is in some tension, but potentially consistent within the range of linear sizes observed for radio galaxies with lower rest-frame peak frequencies.

\subsection{Combined Redshift Constraints}

\subsubsection{\gh}
\label{sec:zcom_host}
Taking the observed line at $1.12\,\mu$m into account and assuming it is one of the stronger lines seen in galaxy spectra, then the two lines most consistent with the bespoke UV-to-radio SED fitting are the [OII]$\lambda 3727$ doublet at $z=2.01$ or Lyman-$\alpha$ at $z=8.21$. These solutions both have high peaks in the {\tt grizli} redshift PDF (Fig.~\ref{fig:grizlizpdf}). We examine the properties of the line and the galaxy in Table~\ref{tab:line} for these two potential redshifts. The full-width half-maximum (FWHM) of the lines comes from the best-fit {\tt grizli} template so should be taken with some caution as should the inferred rest-frame velocity widths. The line fluxes are derived from the normalisation of the line component of the best-fit template. Their differing values can be explained by the larger equivalent width (EW) found for the [OII] solution which is due to the relatively higher continuum model used in that fit. 

\red{We can investigate how consistent the weak features in the grism spectra are with the best fit templates. The Lyman break flux for each of the PEGAS\'E templates for the $z\sim 8$ solution is just under $10^{-14}\,$erg/s/cm$^2$. As a flux density this is equivalent to just under $10^{-19}\,$erg/s/cm$^2$/\AA~ at $1.12\,\mu$m. This value is very close to the observed strength of the feature in Fig.~\ref{fig:0917spec}. The flux of the $4000\,$\AA~ breaks seen in the templates for the $z\sim 2$ solution is around half to one third less, hence is less consistent with the flux observed in the spectrum.} 

We also present in Table~\ref{tab:line} the \red{derived} properties of this radio galaxy for either \red{redshift} option including \red{the} line luminosity, stellar mass, 1.4-GHz radio luminosity as well as constraints on the angular size derived from the rest-frame peak frequency limit \citep[assuming that the source lies on the relation between peak frequency and physical size presented in][]{odea:97}. The limit on the observed peak frequency, $< 50\,$MHz at a $95\%$ confidence, comes from the radio SED fitting in D21 and is held constant in the UV-to-radio fitting (\S~\ref{sec:fittingfull}) . 

For the $z=8.21$ solution, the inferred Lyman-$\alpha$ luminosity is well below the knee of the $z\sim 6.6$ luminosity function \citep{Santos:16} suggesting that the host galaxy does not have an extreme star formation rate or perhaps has moderate extinction in the UV. The stellar mass\red{, from the PEGAS\'E templates, of $10^{10-10.7}\,M_\odot$ is slightly higher than few known galaxies at $8\le z\le 11$ which typically have masses of a few $10^9\,M_\odot$ 
\citep[e.g. A2744\_YD4 at $z=8.38$ and gnz11 at $z=10.96$][respectively]{Laporte:17L,Oesch:16}. However, at this redshift there could be a non-negligible contribution to the rest-frame UV/optical from inverse-Compton (iC) emission either from the radio jet \citep[e.g.][]{ghisellini:14}, or intense compact star-formation, which may lead to an over-estimation of the stellar mass. The inferred limit on the physical size of the radio emission, from the rest-frame turn-over frequency, is consistent with that from the ALMA imaging and the observed interplanetary scintillation at low frequencies radio (as discussed in D21).}

For the $z=2.01$ [OII] solution none of the rest-frame parameters are inconsistent with being at that redshift apart from the size implied by the peak frequency which is mildly inconsistent with the relation from \cite{odea:97}. However, the radio luminosity, $L_{\rm 1.4 GHz}=2.2\times 10^{27}\,$W/Hz, is quite high for a compact, but non-beamed, radio galaxy. Also for this solution one might expect to see [NeV]$\lambda 3426$, [NeVI]$\lambda 3347$ or [MgII]$\lambda 2799$ in the grism spectrum which we do not. 

\begin{table*}[ht]
\begin{threeparttable}
\caption{Derived line and galaxy properties for two different redshift solutions for the $\lambda=1.12\,\mu$m line in the \gh~spectra. These redshifts correspond to known strong emission lines (unlike some solutions in the {\tt grizli} PDF). For both line identifications we present the corresponding redshift, observed-frame wavelength FWHM (from the template), the line flux (from the template fitting), the rest-frame velocity FWHM, the equivalent width (EW), and luminosity of the line for each identification. We also add the stellar mass ($M_{\rm stel}$) from the best-fit {\tt P\'EGASE} templates at that redshift, the $1.4$-GHz radio luminosity from the synchrotron model, the (limit of the) low-frequency rest-frame turn-over (derived from the TPL fitting) and the corresponding constraint on the largest extent in the radio, $\theta_{\rm LAS}$, (from the 0.7\,arcsec angular size) assuming this source follows the size/turn-over relation from \cite{odea:97}.}\label{tab:line}
\begin{tabular}{@{}ccccccccccc@{}}\toprule
line & $z$ & $\Delta\lambda^a$ & Flux & FWHM & EW &  L$_{\rm line}$ & $M_{\rm stel}$ & L$_{\rm 1.4\,GHz}$ & $\nu_{\rm peak}^{\rm rest}$ & $\theta_{\rm LAS}$\\
& &  \AA&erg\,s$^{-1}$cm$^{-2}$&km\,s$^{-1}$ & \AA & erg\,s$^{-1}$ & $\log(M_\odot)$ & W/Hz & GHz & kpc\\
\midrule 
Ly-$\alpha$/1216&8.21&16&$4.7$$\pm$1.8$\times$$10^{-17}$&15&168&$4.4$$\times$$10^{42}$&$10.0-10.7$&$4.3$$\times$$10^{28}$&$\le$$0.50$&$\ge$$1.6$ \\
{\rm [OII]}]$\lambda 3727$ & 2.01&37 & $2.1$$\pm$$0.9$$\times$$10^{-17}$ & 334  & 323 & $4.1$$\times$$10^{41}$ & $10.7$$-$$11.0$     &$2.2$$\times$$10^{27}$&$\le$$0.15$& $\ge$$8.8$\\
\bottomrule
\end{tabular}
\end{threeparttable}
\end{table*}

\subsubsection{Companion Galaxy}

While the bespoke UV-to-radio SED fitting suggests only a broad redshift solution across $0.1<z<3$, when taken in combination with the {\tt EAZY} photometric redshift presented here and the one from D21, then the companion is more likely to be $z\sim 2$. Hence, \red{a probable} identification of the $0.98-\mu$m line is [OII]$\lambda 3727$ at $z=1.63$. \red{The different wavelengths of the weak features seen in each spectrum suggest, despite their close proximity on the sky, that even if the lower redshift solution for \gh~ is correct these two sources likely lie at different redshifts.} 

\subsection{\red{Implications of a Radio Galaxy at $z\sim 2$}}

\red{
Radio galaxies with a luminosity of $L_{\rm 1.4GHz}\sim 4\times 10^{28}\,$W/Hz are not unheard of at $z\sim 2$ \citep[e.g.][]{DeBreuck:10}. However, this source is remarkably compact, smaller than one arcsec, for a source of that luminosity. Most $z\sim 2$ radio galaxies are many tens of kpc in projected extent ($\gg 5\,$arcsec in the observed frame). A few compact sources are comparable in luminosity, e.g. PKS B0008-421 at $z=1.12$ with $L_{\rm 1.4GHz}\sim 4\times 10^{28}\,$W/Hz \citep[][]{Callingham:15,Callingham:17}. PKS B0008-421 turns over at $\sim 1.3\,$GHz in the rest frame and therefore is a gigahertz peaked-spectrum (GPS) source \citep[it is also observed to be $\sim 0.2\,$arcsec in size,][]{King:94}. \gh~would turn over at the much lower rest-frame frequency of $\sim 0.1\,$GHz at $z\sim 2$; hence, from the \citep[][]{odea:97} size-luminosity relation, we may expect it to be larger than the 0.7\,arcsec we measure (see arguments in \S\ref{sec:res_raisered}). The peak of the radio emission could be better constrained by lower-frequency radio observations at $<70\,$MHz, e.g. with LOFAR.}

\red{
From D21, a source at this redshift would have just one CO line (the 3-2 transition at $\sim 115\,$GHz) observable from the spectral observations presented in D21. There is no evidence of a robust line near this frequency although it is at the upper end of the spectral band. Therefore this redshift is adjacent to a gap in our coverage at $1.8\le z\le 2.0$ where we have no constraints on the CO line (see Fig. 6d in D21). If \gh~ was at a redshift of $z=2$, or just above,  the CO luminosity of this source would be limited to $L^\prime_{\rm CO}<$ a few $10^9\,$K\,km\,s$^{-1}$\,pc$^2$, otherwise it is unconstrained.}

\red{
The {\tt RAiSERed} modelling finds that this source would be dominated by an aged lobe if at $z\sim 2$ rather than a young-jet plus aged-lobe model like most compact sources. This hypothesis could easily be tested by future high-resolution VLBI observations. If the radio emission is dominated by more extended synchrotron radiation then the low-frequency turn-over would be more likely due to SSA which is also testable by observations below the MWA frequency band.}

\subsection{Implications of a Radio Galaxy at $z\sim 8$}
\label{sec:disc}

The fitting of the 100-GHz datum suggests a likely contribution from both synchrotron emission and cold dust. If this cold dust were associated with star formation then there could be further iC emission from the compact star-formation. The SED fitting in the far-IR suggests observations at $150-200\,$GHz would be very useful in confirming the presence of cold dust and star-formation. From D21 a source at this redshift may have three CO lines (the 7-6, 8-7 and 9-8 transitions) that fall into the ALMA coverage, but none were detected. This would limit the CO luminosity to $L^\prime_{\rm CO}<$ a few $10^9\,$K\,km\,s$^{-1}$\,pc$^2$.

This result would also confirm our original hypothesis (D20) that such a high-redshift radio galaxy would be compact ($<0.7\,$arcsec) and turn over at low observed-frame frequencies. Very long baseline interferometic (VLBI) radio observations at tens of micro-arcsec would determine the true size of the radio emission. Such observations would help us determine whether the turn-over is due to SSA or FFA. Such observations would also reveal the morphology of this compact source and help test the jet verses lobe hypothesis in \S~\ref{sec:res_raisered}. 

If such a luminous radio source existed in the early Universe it would permit studies of the intervening neutral hydrogen through studies of the absorbed 21-cm line shifted to low frequencies ($154.25\,$MHz). Therefore a sensitive spectral scan from $154\,$MHz to $\sim 189\,$MHz ($\equiv z\sim 6.5$, the approximate end of reionisation) would reveal the distribution of neutral hydrogen along one particular line of sight in the early Universe. Such an observation would likely require the Square Kilometre Array. 

\subsection{\red{Uniqueness of Radio SED}}

\red{
The curvature and steepness in the MWA SED used to select these sources in D20 favours sources which appear to peak below the lowest MWA frequency of 70\,MHz. This curvature is a common characteristic of high redshift radio galaxies. For example, taking the sample of the 158 radio galaxies known at the time to be at $z>2$ from \citet[][]{Miley:08}, we find 76 with GLEAM fluxes of $S_{\rm 150MHz}\ge 0.5\,$Jy. Of these, just under half (34/76) match the mildly updated curvature and steepness criteria for our full sample (Broderick et al., in prep.) - note we do not apply the compactness criteria on the \citet[][]{Miley:08} sample. Above $z=3.5$, most bright radio galaxies (9/11) meet our curvature and steepness criteria.}

\red{
The highest redshift of these, TN J0924$-$2201 at $z=5.19$, has a curved MWA spectrum, but peaks at a higher frequency than the sources meeting our selection criteria. The only other sources similar in luminosity, i.e. $L_{\rm 1.4GHz}>10^{26.5}\,$W/Hz, discovered since are GLEAM\,J0856$+$0223 at $z=5.55$ (D21) and TGSS J1530$+$1049 at $z=5.72$ \citet{Saxena:18b}, both mentioned in the introduction (also see Fig. 1 from Broderick et al., in prep).}
\section{CONCLUSIONS}
\label{sec:con}
We have presented new {\it HST}/WFC3 grism observations of the powerful radio galaxy \gh~ and a companion galaxy $\sim$ one arcsec away. The spectra of these two sources \red{each} show \red{negligible} continuum emission and a faint emission line. These single lines and the overall faintness of these two galaxies makes obtaining their redshift very difficult. Confirmation of faint, potentially very high-redshift sources requires \red{combining} numerous techniques. Hence, we supplement the spectra and the redshift estimat\red{es} by {\tt grizli} with three SED template fitting methods: a bespoke template-plus-synchrotron-function model to the UV-to-radio data, the \red{UV/optical/IR} photometric redshift code {\tt EAZY} and the radio photometric \red{redshift} code {\tt RAiSERed}. Our UV-to-radio SED fitting uniquely combines several key features including the combination of template plus analytical model fitting and proper consideration of upper limits from non-detections. While our results cannot be definitive, we present our best estimate of this source's redshift. From combining the SED fitting as a function of redshift with the faint features in the spectra, our main conclusions are as follows:

\begin{itemize}
    \item For the host galaxy of \gh~the UV-to-radio SED fitting allows for two broad solutions at $z\sim 2$ and $z\sim 8$. Including the $1.12\,\mu$m line seen in the WFC3 spectrum, the host galaxy of GLEAM\,0917$-$0012 most likely lies at either $z=2.01$ or $8.21$. The {\tt EAZY} \red{fitting supports the lower of these two solutions and the} {\tt RAiSERed} fitting supports the higher. \red{The best-fitting UV-to-radio templates at $z\sim 8$, which have a strong Lyman break around $\sim1.1\,\mu$m, are more consistent with the strength of the faint feature seen in the grism spectra than the  best-fitting templates at $z\sim2$.}
    
    \item For the companion the UV to far-IR SED fitting suggests this galaxy is likely at $z<3$ which is consistent with the results from {\tt EAZY} ($z\sim2.6\pm0.5$). Hence, the $0.98\,\mu$m line in the spectrum could well be the O[II]$\lambda 3727$ doublet at $z=1.63$. 

\end{itemize}

However, a categorical determination of the redshift of \gh\ is not yet possible. The best prospects for confirmation lie with either further deep near-IR or millimetre spectroscopy. 

If at $z=8.21$ this galaxy would be the most distant radio-loud AGN known, and well within the EoR. Therefore this source would become a high-priority target for numerous existing and future instruments. Observations with the SKA would allow spectral scans searching for direct and statistical measures of the distribution of neutral hydrogen along the line of sight. Facilities such as {\it JWST} and ground based $30-40\,$m class telescopes could be used to study the host galaxy in the near-IR as well as AGN contributions in the mid-IR. 

\red{If \gh~were to lie at $z\sim 2$ then it would still be one of the most luminous radio-loud AGN known and be extremely compact for it rest-frame low-frequency turn-over. VLBI observations would be key to determine its size and morphology; its radio power is best explained by a lobe-dominated model, but it would be unusually compact for such a source.}

\begin{acknowledgement}
We acknowledge the Noongar people as the traditional owners and custodians of Wadjak boodjar, the land on which the majority of this work was completed. \red{We thank the referee, Gabriel Brammer, for his comments which enhanced the quality of this paper.}

\red{GN acknowledges funding support from the Natural Sciences and Engineering Research Council (NSERC) of Canada through a Discovery Grant and Discovery Accelerator Supplement, and from the Canadian Space Agency through grant 18JWST-GTO1.}

This research is based on observations made with the NASA/ESA {\it Hubble Space Telescope} obtained from the Space Telescope Science Institute, which is operated by the Association of Universities for Research in Astronomy, Inc., under NASA contract NAS 5–26555. These observations are associated with program 16184.

This work has made use of data from the European Space Agency (ESA) mission {\it Gaia} (\url{https://www.cosmos.esa.int/gaia}), processed by the {\it Gaia} Data Processing and Analysis Consortium (DPAC, \url{https://www.cosmos.esa.int/web/gaia/dpac/consortium}). Funding for the DPAC has been provided by national institutions, in particular the institutions participating in the {\it Gaia} Multilateral Agreement.

This publication has made use of data from the VIKING survey from VISTA at the ESO Paranal Observatory, programme ID 179.A-2004. Data processing has been contributed by the VISTA Data Flow System at CASU, Cambridge and WFAU, Edinburgh.

The Hyper Suprime-Cam (HSC) collaboration includes the astronomical communities of Japan and Taiwan, and Princeton University. The HSC instrumentation and software were developed by the National Astronomical Observatory of Japan (NAOJ), the Kavli Institute for the Physics and Mathematics of the Universe (Kavli IPMU), the University of Tokyo, the High Energy Accelerator Research Organization (KEK), the Academia Sinica Institute for Astronomy and Astrophysics in Taiwan (ASIAA), and Princeton University. Funding was contributed by the FIRST program from the Japanese Cabinet Office, the Ministry of Education, Culture, Sports, Science and Technology (MEXT), the Japan Society for the Promotion of Science (JSPS), Japan Science and Technology Agency (JST), the Toray Science Foundation, NAOJ, Kavli IPMU, KEK, ASIAA, and Princeton University. 

This paper makes use of software developed for the Large Synoptic Survey Telescope. We thank the LSST Project for making their code available as free software at  \url{http://dm.lsst.org}.

This paper is based [in part] on data collected at the Subaru Telescope and retrieved from the HSC data archive system, which is operated by the Subaru Telescope and Astronomy Data Center (ADC) at National Astronomical Observatory of Japan. Data analysis was in part carried out with the cooperation of Center for Computational Astrophysics (CfCA), National Astronomical Observatory of Japan. The Subaru Telescope is honored and grateful for the opportunity of observing the Universe from Maunakea, which has the cultural, historical and natural significance in Hawaii. 

This research made use of ds9, a tool for data visualization supported by the Chandra X-ray Science Center (CXC) and the High Energy Astrophysics Science Archive Center (HEASARC) with support from the JWST Mission office at the Space Telescope Science Institute for 3D visualization. This paper also made use of the {\tt TOPCAT} software \citep{Taylor:05}. 

JA acknowledges financial support from the Science and Technology Foundation (FCT, Portugal) through research grants PTDC/FIS-AST/29245/2017, UIDB/04434/2020 and UIDP/04434/2020.

\end{acknowledgement}

%% file: tables/flux.tex
$g_{\rm~HSC}$  &  0.48  &  27.0  &  0.77  &  1.506  &  2.78  & $ 0.0509 \pm 0.0407 $ & $ 0.576 \pm 0.0703 $ \\
$r_{\rm~HSC}$  &  0.62  &  27.0  &  0.58  &  1.703  &  2.64  & $ 0.214 \pm 0.0825 $ & $ 0.790 \pm 0.112 $ \\
$i_{\rm~HSC}$  &  0.77  &  27.0  &  0.76  &  1.332  &  2.65  & $ 0.151 \pm 0.0553 $ & $ 0.760 \pm 0.0927 $ \\
$z_{\rm~HSC}$  &  0.89  &  27.0  &  0.68  &  1.567  &  2.62  & $ 0.143 \pm 0.150 $ & $ 0.770 \pm 0.168 $ \\
$y_{\rm~HSC}$  &  0.97  &  27.0  &  0.68  &  1.553  &  2.67  & $ 0.309 \pm 0.203 $ & $ 0.824 \pm 0.216 $ \\
F098M  &  0.98  &  25.666  &  0.083  &  1.089  &  2.03  & $ 0.292 \pm 0.144 $ & $ 0.545 \pm 0.151 $ \\
F105W  &  1.05  &  26.265  &  0.092  &  1.093  &  1.14  & $ 0.429 \pm 0.0487 $ & $ 0.871 \pm 0.0901 $ \\
$J_{\rm~VIK}$  &  1.25  &  30.0  &  0.85  &  1.2  &  19.8  & $ -0.0670 \pm 3.48 $ & $ 0.475 \pm 3.56 $ \\
$H_{\rm~VIK}$  &  1.6  &  30.0  &  0.85  &  1.2  &  19.4  & $ -1.16 \pm 5.85 $ & $ 1.19 \pm 6.07 $ \\
$K_s$(HAWKI)  &  2.15  &  30.152  &  0.8  &  1.327  &  0.666  & $ 3.17 \pm 0.346 $ & $ 1.56 \pm 0.205 $ \\